%% file: Paper.tex
\documentclass[journal,onecolumn,12pt]{IEEEtran}

\usepackage{multirow}
\usepackage{graphicx}
\usepackage{color}
\usepackage{subfigure}
\usepackage{epsfig}
\usepackage{citesort}
\usepackage{setspace}
\usepackage{latexsym}
\usepackage{amssymb}

\title{Design and Characterization of a Full-duplex Multi-antenna System for WiFi networks}
\author{Melissa Duarte$^\diamond$, Ashutosh Sabharwal$^\diamond$, Vaneet Aggarwal$^\dagger$, Rittwik Jana$^\dagger$, K.~K.~Ramakrishnan$^\dagger$, Christopher Rice$^\dagger$ and N.~K.~Shankaranarayanan$^\dagger$\thanks{$^\diamond$Department of Electrical and Computer Engineering, Rice University, Houston, TX 77005. }\thanks{$^\dagger$AT\&T Labs-Research, Florham Park, NJ 07932.} }
\newtheorem{result}{Result}

\begin{document}
\maketitle
\input{abstract}
\input{intro2}

\input{design}

\input{expt2}

\input{mac_v}

\input{macsim2}

\input{concl}

{
\bibliographystyle{IEEEbib}
\bibliography{fdbib}
}

\input{apdmac}
\input{figtab}

\end{document}

%% file: abstract.tex
\begin{abstract}

In this paper, we present an  experimental and simulation based study to evaluate the use of full-duplex as a mode in practical IEEE~802.11 networks. To enable the study, we designed a 20~MHz multi-antenna OFDM full-duplex physical layer and a full-duplex capable MAC  protocol which is backward compatible with current 802.11.  Our extensive over-the-air experiments, simulations and analysis demonstrate the following two results. First, the use of multiple antennas at the physical layer leads to a higher ergodic throughput than its hardware-equivalent multi-antenna half-duplex counterparts, for SNRs above the median SNR encountered in practical WiFi deployments. Second, the proposed MAC translates the physical layer rate gain into near doubling of throughput for multi-node single-AP networks. The two combined results allow us to conclude that there are potentially significant benefits gained from including a full-duplex mode in future WiFi standards.

\end{abstract}

%% file: intro2.tex
\newpage
\section{Introduction}
\label{sec:intro}

Currently deployed wireless communications systems cannot transmit and receive on the same frequency band at the same time, i.e.,\ networks do not operate in a single-channel full-duplex fashion. As a result, networks are either time-division duplex (e.g.,\ WiFi) or frequency-division duplex (e.g.,\ cellular). The key challenge in achieving true full-duplex communication is the large power differential between the ``self-interference" created by a node's own radio transmission and the signal of interest originating from a distant node. The large power differential is simply because the self-interference signal has to travel much shorter distances compared to the signal of interest. As a result of the large power differential, the signal of interest is swamped by the self-interference in digital baseband due to finite resolution of analog-to-digital conversion.

Full-duplex experimental demonstration for narrowband systems was first reported in 1998~\cite{Chen:1998aa}. Since then, multiple authors~\cite{Khandani,Bliss:2007,Radunovic:2009aa,Choi,Duarte,Jain,achalarxiv,achal,Khojastepour,Aryafar} have reported different methods and implementations for various single and multiple antenna extensions.
However, till date none of the prior methods  have reported experimental evidence to achieve long-enough communication ranges (best reported number in all prior literature is 8~meters with line-of-sight)
for full-duplex to be considered in WiFi-like systems. Our focus in this paper is to investigate if a practical WiFi system can leverage full-duplex gains for its typical communication range. Our experiment based analysis is the first to investigate the performance of full-duplex systems over the entire range of signal to noise ratio (SNR) values typical in WiFi communications.

In this paper, we present  a multi-antenna wideband PHY and MAC design to enable a practical full-duplex mode in WiFi. Via extensive over-the-air tests, we show that our design achieves higher throughputs than its hardware-equivalent half-duplex MIMO counterparts, for a significant portion of the WiFi communication range. Our contributions in design are two-fold. First, to reduce the self-interference, the PHY uses a combination of three methods: (i)~passive suppression via appropriate placement of multiple antennas on a device, (ii)~a per-subcarrier per-receive-antenna analog self-interference canceler for MIMO OFDM systems and (iii)~a digital self-interference canceler implemented in baseband. Second, to gauge realistic gains in actual systems, the MAC design leverages legacy WiFi RTS/CTS packets to seamlessly support legacy half-duplex and new full-duplex  modes. By design, the MAC is minimally different from IEEE~802.11 and is designed to leverage the existing 802.11 ecosystem to accelerate potential adoption of the full-duplex mode in future 802.11 revisions.

En route to showing that our design provides rate gains for WiFi networks, we perform extensive statistical characterization of the design elements, revealing several new findings.  Our findings can be divided into three categories: (i)~self-interference canceler performance in full-duplex, (ii)~the comparison of empirical ergodic rates achieved by full- and half-duplex systems, and (iii)~an extensive MAC layer performance analysis for different traffic scenarios.

{\bf Results on self-interference cancellation in full-duplex}: Recall that we are employing three mechanisms to reduce self-interference -- passive suppression by antenna placement, and two active cancelers -- one in analog and the other in digital baseband. The three mechanisms are concatenated serially to result in a three-stage design. The serial concatenation implies that each stage is operating on the residual signal of the previous stage. As a result, the performance of each stage is \emph{not} independent of the performance of the stages prior to it. In general, if a stage cancels more of the self-interference, then the subsequent stages cancel less. Thus, in general, our results show that the total self-interference canceled by any two concatenated stages is \emph{not} the sum of maximum self-interference canceled by each stage individually in isolation. The non-additive nature of concatenated cancellation techniques also demonstrates the challenge of completely suppressing self-interference -- individually improving each stage does not guarantee equivalently better performance in the total system performance.

Digging deeper to understand the role and interaction of each cancellation stage, we show  following four results experimentally for different 20~MHz 64-subcarrier OFDM physical layers.  First, we consider antenna placement for 2$\times$1 MISO full-duplex, where each node has three antennas -- two transmit and one receive. By placing antennas \emph{around} the device to use the device itself to attenuate self-interference and also leveraging antenna polarization, self-interference can be suppressed by an additional 15~dB compared to the configuration where there is no device. Thus, the key message is that  placement of antennas is crucial in full-duplex devices. We note that our antenna placement aims to only increase the pathloss of self-interference and thus is highly robust to device size variations. In contrast,  prior antenna placement techniques aim to create beamforming nulls~\cite{Choi,Khojastepour,Aryafar}, which are designed under the assumption that the self-interference channel does not have multi-path components. As a result, beam-forming based designs in~\cite{Choi,Khojastepour,Aryafar} require self-interfering antennas to be either symmetrically spaced or placed at a distance which is a function of the frequency of operation.

Second, passive device based suppression largely reduces the direct line-of-sight path for self-interference and thus the multi-path reflections become dominant. This becomes evident by the fact that with more passive cancellation, the self-interference channel becomes more frequency-selective. The measured frequency-selectivity was our motivation behind per-subcarrier analog canceler, which actively cancels self-interference in each OFDM band with sub-band specific cancellation coefficients.

Third, we measure the performance of each cancellation stage and also consider its impact on subsequent stages. The measured results clearly show the above mentioned fact -- more cancellation by one stage means lower cancellation possible in later stages. In~\cite{Choi}, it was assumed that the performance of cancellation stages is additive. This assumption was then used to obtain an estimate of 73~dB of total analog plus digital cancellation. However, the prototype implemented in \cite{Choi} achieved 30~dB of analog plus digital cancellation, which is 43~dB less than their estimated maximum of 73~dB. Hence, the total cancellation of concatenated cancellation stages did not equal the sum of the cancellation achieved by each stage individually in isolation. Thus, we believe that our conclusions are qualitatively typical for any hardware implementation, which uses serial concatenation of different cancellation schemes.

Lastly, we combine all three methods of cancellation (passive, analog, and digital) and demonstrate that our three-stage self-cancellation system achieves a median cancellation of 85~dB, with minimum of 70~dB and a maximum of 100~dB. The median and maximum numbers for \emph{total} self-interference cancellation are the best reported numbers in the literature till date. We note the importance of studying the statistical properties of the cancelers. All cancellation mechanisms rely on some form of channel estimation to adjust its cancellation weights and thus have to deal with noise introduced by RF and baseband (e.g. in the form of quantization noise) stages. Thus, no cancellation mechanism can be guaranteed to achieve a constant cancellation in all cases, and will  exhibit statistical variations.

{\bf Results on Ergodic rate Comparisons}: We implemented two full-duplex physical layers -- 2$\times$1 MISO and 1$\times$1 SISO, and three half-duplex systems -- 2$\times$1 MISO, 3$\times$1 MISO and 2$\times$2 MIMO. The RF hardware usage of the five systems is compared by counting the \emph{total} number of RF up-conversion and down-conversion chains. A 2$\times$2 MIMO half-duplex uses 2 up-converting and 2 down-converting chains for a total of 4 chains. A 2$\times$1 MISO full-duplex uses 3 up-converting and 1 down-converting chains, again for a total of 4 chains. Similarly all other configurations mentioned above use 4 or fewer total chains. The main motivation for using RF hardware equivalence is that in most portable devices, the power consumption of RF is a key factor and thus often determines the largest supported antenna configurations.

We perform extensive experiments which allow us to compare the performance of full-duplex and half-duplex systems for SNR values from 0 to 40~dB. In WiFi systems, the received signal of interest power is typically between $-80$~dBm and $-60$~dBm, and the noise floor is around $-90$~dBm. Hence, the range of operation for WiFi systems corresponds to SNR values lower than 30dB. We observe that for a significant fraction of the WiFi SNR range of operation (more specifically, SNR values greater than 20 dB), 2$\times$1 full-duplex can often outperform the rest of the four configurations. In terms of multiplexing gain, 2$\times$1 full-duplex and  2$\times$2 half-duplex should have the same multiplexing gain of two. However, the measured multiplexing gain of 2$\times$2 half-duplex is often less than two, and here again 2$\times$1 full-duplex achieves a higher measured multiplexing gain. While surprising at first, the result is easily explained by the distribution of condition numbers of channel matrices observed in our extensive indoor tests,\footnote{We performed only indoor tests since most WiFi deployments are indoors.} and match the results for half-duplex MIMO systems observed in other experiments~\cite{cond}.

{\bf Results on  MAC Layer}: Recall that a primary design objective for the MAC was to make minimal changes to legacy 802.11 MAC to extract the advantages of full-duplex technology. The  MAC design supports both legacy half-duplex  and full-duplex flows without hurting the throughput for half-duplex nodes significantly. For the full-duplex flows, asymmetric packet sizes are also supported since the packet sizes in the two directions of a full-duplex transfer may be different.

In 802.11, if a node cannot decode a frame successfully, it triggers a longer wait time due to the use of EIFS (Extended Inter Frame Spacing). In the full-duplex mode, the nodes other than the two nodes participating in full-duplex exchange, do not decode frames correctly. Thus the  MAC design has modification to avoid waiting for EIFS in certain scenarios. The new MAC design is simulated in an OPNET based MAC simulator, which allowed us to use an industry-standard WiFi implementation and stay backward compatible with IEEE~802.11 MAC protocol. We focused on the full-buffer scenario to determine maximum throughput, and also examined fairness between the full-duplex and half-duplex nodes.

Our results were obtained in four major steps. First, we evaluated the performance of a single AP communicating with one full-duplex flow.
Full-Duplex MAC doubles the throughput of the system as compared to a legacy WiFi half-duplex communication using RTS/CTS signaling for a fixed total cancellation of 85 dB, propagation loss of 63dB, and symmetric traffic. Full-duplex MAC throughput increases by 87\% as compared to legacy WiFi half-duplex system that does not use RTS/CTS signalling for packet size of 1500 bytes. We further investigated asymmetric packet sizes, where uplink data packet size can be different from downlink packet size for a full-duplex exchange. Keeping in mind that typical data communications  uses TCP as the transport layer protocol, in which a one-way transfer of data would typically have 1500 byte data packets with 40 byte acknowledgment in the reverse direction, we quantify the goodput performance for varying packet sizes. As the degree of asymmetry reduces, the throughput gains ranged from 30\%--100\% as compared to a legacy half-duplex system \emph{with} RTS/CTS, and 18\%--87\% as compared to a legacy half-duplex system \emph{without} RTS/CTS.

Second, we considered scaling of full-duplex system as one AP communicates with more full-duplex nodes. We first note that for a half-duplex system with four or more nodes, the use of RTS/CTS improves the goodput since data collisions (which trigger retransmission of the large data packets) are replaced by collisions of the short RTS frames. We find that the sum throughput for a full-duplex system increases by a factor of at-least two, when compared to a half-duplex system with RTS/CTS with the same number of nodes. 

Third, we consider the system performance when full-duplex nodes co-exist with half-duplex nodes, where the half-duplex nodes' MAC logic has been slightly modified to ignore collisions during a full-duplex transfer. This provides an insight into the dynamics of co-existence. This modification in half-duplex can be made by a shift in logic of timing without the need for new hardware and may or may not be pragmatic.
 We find that for a system with $m$ full-duplex and $m$ half-duplex nodes, the total throughput compared to a  half-duplex-only system increases by a factor of $1+m/(2m+1)$. The percentage increase in throughput for a mixed system as compared to a half-duplex system increases with $m$ such that the maximum percentage increase can be up to 50\%. The uplink and downlink throughputs of the full-duplex nodes in a mixed system are higher as compared to the uplink and downlink throughputs of a node in a system with $2m$ half-duplex nodes and no full-duplex node respectively. Thus, the improved hardware for full-duplex nodes provides a substantial improvement to the throughput of the full-duplex nodes. In addition, half-duplex nodes also achieve higher throughputs. The downlink throughput from AP to half-duplex nodes almost doubles when there is a mix of full- and half-duplex nodes, and the uplink throughput from half-duplex node to AP is also improved slightly as compared to the corresponding throughput in a purely half-duplex system.

Finally, we consider the coexistence with legacy half-duplex nodes that have no modifications.  Much like the above discussion, the uplink and downlink throughputs of full- duplex nodes increases, and so does the downlink throughput to HD nodes as compared to a purely half-duplex system. However, half-duplex nodes do not grab the channel as often as they would in a purely half-duplex system (where the total number of nodes in two cases are the same) leading to a decrease in their uplink throughput by around 40\% for $m=2$ as compared to purely half-duplex system. To increase probability of access, we change the proposed full-duplex MAC design to make it better throughput fair with the legacy half-duplex nodes. The above change in full-duplex MAC decreases throughput by around 2\% for the full-duplex nodes (as compared to the case of modified half-duplex and unchanged full-duplex nodes) by making them less aggressive in lieu of increased probability of access for legacy half-duplex nodes which is almost the same as if all the nodes were half-duplex.

The rest of the paper is organized as follows. In Section II, we describe the MIMO wideband canceller design which uses a combination of passive suppression and active cancellation techniques. In Section III, we describe the experimental setup for validating the design. Section IV and V evaluates the cancellation design in terms of cancellation and throughput respectively. In Section VI, we give our MAC design with detailed evaluations in Section VII. Section VIII concludes this paper.

%% file: design.tex
\section{MIMO Wideband Canceller design}
\label{design}

We present a design for a wideband multiple antenna self-interference canceller which uses a combination of passive suppression and active cancellation techniques, where passive suppression precedes active cancellation.  The cancellation techniques are explained below.


{\bf Passive Suppression (PS):} Passive suppression is achieved by maximizing the attenuation of the self-interference signal due to propagation path loss over the self-interference channel, which is the channel between same node transmitter and receiver antennas. The amount of passive suppression depends on the distance between antennas, the antenna directionality, and the antenna placement on the full-duplex device. We use $\mathbf{h}_{i,m,n}$ to denote the self-interference channel between transmitter antenna $m$ and receiver antenna $n$ at node $i$. The self-interference channel, $\mathbf{h}_{i,m,n}$, varies with time and frequency due to changes in the node's environment. Our design of self-interference cancellation for OFDM systems will be presented in the frequency domain. We use $h_{i,m,n}[k]$ to denote the magnitude and phase that the self-interference channel $\mathbf{h}_{i,m,n}$ applies to subcarrier $k$. For a system with $K$ subcarriers the channel vector is defined as $\mathbf{h}_{i,m,n}=[h_{i,m,n}[1],h_{i,m,n}[2], \cdots, h_{i,m,n}[K]]$. Figure~\ref{fig:fdnode} shows the two passive cancellation paths $\mathbf{h}_{i,1,1}$ and $\mathbf{h}_{i,2,1}$ for a full-duplex node with two transmitter antennas and one receiver antenna.


{\bf Active Analog Cancellation (AC):} As the name suggests, the active cancellation is performed in analog domain \emph{before} the received signal passes through the Analog-to-Digital Converter (ADC). For an OFDM MIMO node, the self-interference signal received at Node $i$ antenna $n$ on subcarrier $k$ after passive suppression is equal to $y_{i,n}^{PS}[k] = \sum_{m=1}^{M}h_{i,m,n}[k]x_{i,m}[k]$, where $x_{i,m}[k]$ is the signal transmitted from Node $i$ on subcarrier $k$ antenna $m$. Analog cancellation of the self-interference at receiver antenna $n$ is implemented by subtracting an estimate of $y_{i,n}^{PS}[k]$ from the received signal.

In our proposed MIMO wideband canceller design, the additional hardware components required for active analog cancellation of the self-interference at one receiver antenna consist of one Digital-to-Analog converter (DAC), one up-converting radio chain (Tx Radio) which up converts the signal from Base Band (BB) to Radio Frequency (RF), one fixed attenuator, and one RF adder. Figure~\ref{fig:fdnode} shows a diagram of our proposed analog cancellation for a full-duplex node with two transmitter antennas and one receiver antenna. One input to the RF adder is the signal at the receiver antenna, and the other input is a canceling signal $\mathbf{z}_{i,n}$ local to node $i$ which is input to the RF adder via a wire. For subcarrier $k$ and receiver antenna $n$, the local signal $\mathbf{z}_{i,n}$ is equal to $z_{i,n}[k]=-h_{i,n}^{W}[k]\sum_{m=1}^{M}b_{i,m,n}[k]x_{i,m}[k]$, where $h_{i,n}^{W}[k]$ denotes the magnitude and phase that affect a signal at subcarrier $k$ when passing through the wire connected to the RF adder at node $i$ receiver antenna $n$. Further, $b_{i,m,n}[k]$ denotes the cancellation coefficient for the self-interference received at antenna $n$ from transmitter antenna $m$ at subcarrier $k$ at Node $i$.


The self-interference at subcarrier $k$ after analog cancellation at antenna $n$  (this is the signal at the output of the RF adder connected to antenna $n$) is equal to $y_{i,n}^{AC}[k]$$=y_{i,n}^{PS}[k] -z_{i,n}[k]$, which can be rewritten as $y_{i,n}^{AC}[k]$$=\sum_{m=1}^{M}(h_{i,m,n}[k]-h_{i,n}^{W}[k]b_{i,m,n}[k])x_{i,m}[k]$. From the equation for $y_{i,n}^{AC}[k]$, we observe that active analog cancellation achieves perfect cancellation when $b_{i,m,n}[k]=h_{i,m,n}[k]/h_{i,n}^{W}[k]$. In a real system, $h_{i,m,n}[k]$ and $h_{i,n}^{W}[k]$ can only be estimated, which leads to the following  computation of
\begin{equation}
b_{i,m,n}[k]=\widehat{h}_{i,m,n}[k]/\widehat{h}_{i,n}^{W}[k],
\label{eqn:cancoeff}
\end{equation}
where $\widehat{h}_{i,m,n}[k]$ and $\widehat{h}_{i,n}^{W}[k]$ are the estimates of $h_{i,m,n}[k]$ and $h_{i,n}^{W}[k]$ respectively. Thus, cancellation is usually not perfect. The estimates of $h_{i,m,n}[k]$ and $h_{i,n}^{W}[k]$ are computed based on pilots sent from each transmitter radio on orthogonal time slots.

In a WiFi system that uses RTS/CTS, the estimates of $h_{i,m,n}[k]$ and $h_{i,n}^{W}[k]$ can be computed based on pilots sent during the RTS/CTS transmissions. Further, since $h_{i,n}^{W}[k]$ is a wire, it is a static channel and it does not need to be estimated often. While the RTS/CTS packet exchange adds overhead to the system, it enables full-duplex and results in overall rate gains as will be shown in Sections~\ref{sec:mac} and~\ref{macsim}.


We note that any additional transmitter radio used for analog cancellation does not require a power amplifier since it is transmitting over a wire. However, for our specific implementation, the radio used for analog cancellation had a power amplifier which could not be removed. Hence, we used a fixed RF attenuator connected in series, as shown in Figure~\ref{fig:fdnode}, in order to reduce the signal power levels at the output of the canceller radio to the levels required for cancellation. The attenuator used was a passive device (part number PE7001~\cite{PE7001data}) that attenuates all the frequencies in the band of interest by the same amount. The value set for the attenuator was a function of the antenna configuration used because different antenna configurations resulted in different levels of self-interference power at the receiver antenna; different antenna configurations have different amount of passive suppression as will be shown in Section~\ref{subsec:phy_canceller:passive}. The four antenna configurations used are shown in Table~\ref{tbl:antenna_config} and will be explained in more detail in Section~\ref{subsec:expsetup:antennaplacements}. The attenuator was set equal to 35~dB for Antenna Placement 1 without device, 45~dB for Antenna Placement 1 with device, 50~dB for Antenna Placement 2 without device, and 55~dB for Antenna Placement 2 with device. The RF attenuator would not have been needed if the radio used for analog cancellation had a larger range of output powers and did not use a power amplifier by default.

We highlight that the RF adder used for analog cancellation is a passive device (part number PE2014~\cite{PE2014data}) and applies the same addition operation to all the frequencies in the band of interest.



{\bf Digital Cancellation (DC):} There is a residual self-interference $y_{i,n}^{AC}[k]$ that remains after analog cancellation due to imperfect analog cancellation. Active digital cancellation estimates $y_{i,n}^{AC}[k]$ and subtracts this estimate from the received signal in the digital domain. The estimate of $y_{i,n}^{AC}[k]$ is computed based on a second round of pilots sent from each transmitter antenna and received while applying analog cancellation to each receiver antenna. Specifically, the second round of pilots is used to compute $h_{i,m,n}[k]-h_{i,n}^{W}[k]b_{i,m,n}[k]$. Alternatively, the estimate of  $y_{i,n}^{AC}[k]$ can be computed without extra pilots if implemented based on correlation between the transmitted and received self-interference payload signal.

%% file: expt2.tex
\section{PHY Experiment Description}
\label{sec:expsetup}

In this section, we describe our experiment testbed, antenna configurations and physical layer techniques which will be compared and their implementation details on WARP~\cite{WARP}.

\subsection{Node Locations}
\label{subsec:expsetup:locations}

We used five nodes, labeled as nodes Na, Nb, Nc, Nd, and Ne. The nodes were placed at locations shown in Figure~\ref{setup_loc}. Nodes Na, Nb, Nc, Nd, and Ne were located at a height of 1.5~m, 1.5~m, 1.4~m, 1.7~m and 2.0~m respectively, above the floor. Experiments were conducted in the second floor of a three-floor office building and were performed both at night and during office work hours with people walking in and out of the rooms. The five-node setup allowed us to evaluate ten different two-node links. The ten link pairs, their inter-node distance and the type of channel for each link are shown in Table~\ref{tbl:linkdistance}. Our choices allowed us to create line-of-sight channels and also extremely challenging multi-wall propagation environments, which represented a typical Wi-Fi deployment. In contrast, the experiment setup in~\cite{Aryafar} was located at least 20 m from the the nearest wall hence, which does not capture some typical WiFi scenarios. For experiment results in~\cite{Jain,Choi,Aryafar} the distance between communicating nodes was not reported.


\subsection{Full-duplex and Half-duplex Modes}
\label{subsec:expsetup:fdhdconsidered}

For each of the ten links, we ran experiments for the following physical layers: full-duplex 1$\times$1~(FD1$\times$1), full-duplex 2$\times$1~(FD2$\times$1), half-duplex 2$\times$1~(HD2$\times$1), half-duplex 3$\times$1~(HD3$\times$1), and half-duplex 2$\times$2 (HD2$\times$2). Experiment results obtained for the above five systems have the necessary data to evaluate the performance of our full-duplex  design and compare its performance with half-duplex systems which use the same or less radio resources per node. Notice that an HD$M\times N$ node needs $M$ up-converting radio chains and $N$ down-converting radio chains for a total of $M+N$ radio chains. In contrast, our proposed FD$M'\times N'$ node uses $M'$ up-converting radio chains for transmission, $N'$ down-converting radio chains and $N'$ up-converting radio chains for self-interference cancellation for a total $M'+2N'$ radio chains per node for any $M', N' \ge 1$.

For all five PHY configurations listed above, the total number of chains is no more than 4. That is $M+N \leq 4$ for half-duplex systems and $M'+2N' \leq 4$ for full-duplex systems.
Table~\ref{tbl:numradios} shows the number of radios and antennas per node used by each of the full-duplex and half-duplex systems considered. We will compare the performance of full-duplex and half-duplex systems which use the same number of radios per node. The performance of FD2$\times$1 will be compared with the performance of HD3$\times$1 and HD2$\times$2 systems. The performance of FD1$\times$1 will be compared with the performance of HD2$\times$1.

For the experiments with more than one transmitter antenna, the multiple antenna codes used were the following. For the FD2$\times$1 experiments we used an Alamouti code~\cite{alamouti}. Hence, in Figure~\ref{fig:fdnode}, the signals $x_{i,1}$ and $x_{i,2}$ correspond to Alamouti encoded symbols. The HD2$\times$1 experiments also used an Alamouti code. The HD3$\times$1 experiments used a rate 3/4 orthogonal space-time block code (OSTBC) from MATLAB MIMO library~\cite{matlab}. The HD2$\times$2 experiments used spatial multiplexing for two spatial streams and the receive processing was implemented using channel inversion.

We note that a FD2$\times$1 Alamouti implementation using our proposed wideband MIMO canceller requires three antennas per node which is less than what is required by the MIMO cancellation techniques proposed in~\cite{Aryafar}. The MIMO antenna cancellation technique in~\cite{Aryafar} would require at least four antennas per node. The transmitter/receiver antenna cancellation technique proposed in~\cite{Aryafar} requires 6 antennas per node for implementation of a FD2$\times$1 Alamouti system.


\subsection{Multi-antenna Placements}
\label{subsec:expsetup:antennaplacements}

We considered two possible antenna placements for the full-duplex and half-duplex experiments. For each antenna placement we considered two cases: antennas with a device (a 15-inch Macbook Pro laptop) and without a device. Hence, we considered a total of four different configurations as shown in Table~\ref{tbl:antenna_config}. For all the configurations, R1 was used as the receive antenna for all the systems that used only one receiver antenna, i.e FD1$\times$1, FD2$\times$1, HD2$\times$1, and HD3$\times$1. For HD2$\times$2, all the configurations used R1 and R2 as receiver antennas. For all the configurations and systems evaluated, if $M$ antennas were required for transmission, we used antennas $T1$ to $TM$.





The antennas used in experiments~\cite{Lcom} are designed for 2.4 GHz operation, with vertical polarization, and have toroid-like radiation pattern shown in~\cite{Lcom}. In Antenna Placement~1 (A1), the full-duplex experiments correspond to the case where the main lobe of the receiver antenna (R1) is in the same direction as the main lobe of T1 and orthogonal to the main lobe of T2. In Antenna Placement~2 (A2), the full-duplex experiments correspond to the case where the receiver (R1) main lobe is orthogonal to the main lobe of both T1 and T2. As experiments will demonstrate, the orthogonal placement of the transmitter and receiver main lobes in A2 will help reduce the self-interference. Hence A2 will result in larger passive suppression than A1. Experiment results in Section~\ref{sec:phy_canceller} will also demonstrate and quantify the increase in passive suppression achieved by placing  antennas appropriately around a device.

\subsection{Transmit Power Normalization}
\label{subsec:expsetup:txp}

For a fair comparison between full-duplex and half-duplex systems, the total energy transmitted by a full-duplex node must be the same as the total energy transmitted by a half-duplex node. Since energy is power times transmission time, the equation 
\begin{equation}
P_i^{FD}T_i^{FD}=P_i^{HD}T_i^{HD}
\label{eqn:powers}
\end{equation}
defines the relationship between full- and half-duplex powers, where $P_{i}^{FD}$ denotes the transmission power use by Node $i$ in full-duplex mode, $P_{i}^{HD}$ denotes the transmission power used by node $i$ in half-duplex mode, $T_{i}^{FD}$ denotes the duration of a transmission from node $i$ in full-duplex mode, and $T_{i}^{HD}$ denotes the duration of a transmission from node $i$ in half-duplex mode.

Consider a finite duration, $\tau$, of time for bi-directional communication between Nodes~1 and~2.  From time constraints for full-duplex and half-duplex we have that $T_{1}^{FD}=T_{2}^{FD}=\tau$ and $T_{1}^{HD}+T_{2}^{HD}=\tau$.  We define $\beta=T_{1}^{HD}/\tau$. Using Eq. (\ref{eqn:powers}), the definition of $\beta$, and the time constraints, we obtain that for a fair comparison between full-duplex and half duplex systems the node powers used in full-duplex and half-duplex must satisfy
\begin{eqnarray}
P_{1}^{FD}&=&P_{1}^{HD}\beta \label{eqn:P1FD} \\
P_{2}^{FD}&=&P_{2}^{HD}(1-\beta). \label{eqn:P2FD}
\end{eqnarray}

Notice that Equations~(\ref{eqn:P1FD}) and (\ref{eqn:P2FD}) do not impose any constraint on the maximum power assigned to a node. However, in real systems, the maximum instantaneous radiated power is limited and is typically defined in standards. Hence, in order to include practical considerations in our power assignment equations, we define $\Pi$ as the maximum power that can be radiated by the network (not just one node, but all the nodes in the network together) at any time. Since half-duplex transmissions from each node are orthogonal in time, it implies that in a network with two nodes $i=1,2$ the transmission powers must be such that $P_{1}^{HD}\leq\Pi$ and $P_{2}^{HD}\leq\Pi$. In contrast, since full-duplex transmissions from each node are simultaneous, the instantaneous radiated power constraint of $\Pi$ translates to a power constraint of $P_{1}^{FD}+P_{2}^{FD}\leq\Pi$ for full-duplex nodes. Thus, we ensure that at any given time, a network with full-duplex nodes radiates the same power that would be radiated by a network with half-duplex nodes.

All our experiments correspond to an instantaneous power constraint of $\Pi=8$ dBm and we achieve this constraint with equality. Hence, our experiments correspond to tŒhe following transmit power assignments:  $P_{1}^{HD}=P_{2}^{HD}=8$ dBm, $P_{1}^{FD}=8\mbox{ dBm} + 10 \log_{10}(\beta)$, and $P_{2}^{FD}=8\mbox{ dBm} + 10 \log_{10}(1-\beta)$. We performed only symmetric experiments, where $\beta=0.5$, leading to $P_{1}^{FD}=P_{2}^{FD}=5$ dBm.

The radios used in our experiments can transmit at a maximum power of 25~dBm. However, we observed that the radio's transmitter power versus gain setting relation is linear only for output powers between 0 dBm and 15 dBm for OFDM signals of 20 MHz bandwidth used in our experiments. Consequently, for our experiments we chose transmission powers which lie close to the middle of the linear range of the transmitter radios. Accounting for amplifier nonlinearities and their impact on cancellation coefficients, $b_{i,m,n}$, will be focus of future work.

\subsection{WARP Implementation and Testbed Setup}
\label{subsec:expsetup:implementation}

The digital and analog signal processing at a node were implemented using the WARPLab framework~\cite{WARP}. The WARPLab framework facilitates experiment implementation by allowing the use of MATLAB for digital signal processing and the use of WARP~\cite{WARP} hardware for real-time over-the-air transmission and reception.

All full-duplex and half-duplex experiments were conducted at a 2.4 GHz Wi-Fi channel without any other concurrent traffic. In all our experiments the nodes shared the same carrier frequency reference clock.
All systems implemented have a bandwidth of 20 MHz using 64 subcarriers with 48 subcarriers used for payload as specified in one of the possible Wi-Fi modes.

For each of the ten links considered, we ran experiments with both nodes using the same antenna/device configuration, and we considered all the possible combinations for the ten different links and four possible configurations shown in Table~\ref{tbl:antenna_config}. Thus there were a total of 40 different scenarios. For each scenario and full-duplex/half-duplex system considered, an experiment consisted  of transmitting 90 packets from each of the nodes in the link. Each packet transmitted consisted of 68 OFDM symbols (the number of OFDM symbols per packet was limited by buffer sizes in the WARPLab framework) and each subcarrier was modulated using QPSK. Since there were 48 payload subcarriers per OFDM symbol, the total number of bits transmitted per packet per node was equal to 6528 and the total number of bits transmitted per node in 90 packets was equal to 587,520.

\section{PHY Evaluation: Canceller Performance}
\label{sec:phy_canceller}

In this section we characterize the performance of the self-interference cancellation stages. We demonstrate that our full-duplex  design can achieve self-interference cancellation values, which can be larger than what has been reported in prior work.

\subsection{Metric for Canceller Analysis}
\label{subsec:phy_canceller:measurement}

We measured the self-interference power after each stage of cancellation for each packet transmitted by a node in full-duplex mode. For each stage of cancellation, the amount of cancellation (in dB) was computed as the difference between the self-interference power before cancellation and the self-interference power after cancellation. The measurement of the self-interference power after each cancellation stage was computed based on the RSSI reading provided by the WARP radios.  A more detailed explanation of the power measurements is provided in~\cite{MelissaThesis}.


\subsection{Performance of Passive Suppression}
\label{subsec:phy_canceller:passive}


\begin{result}[Gain from Antenna Placement and Orientation]
\emph{The amount of passive suppression increases, by as much as 15~dB, for the placement where (a)~the receiver antenna is placed orthogonal to the transmitter antennas responsible for self-interference and (b)~the device-induced pathloss is increased.}
\label{result:cancellation:1}
\end{result}

Figure~\ref{passive} shows a characterization of the amount of passive suppression achieved by the four different configurations listed in Table~\ref{tbl:antenna_config}. First, we observe that at a CDF value of 0.5, configuration A1 with device achieves approximately 10 dB better cancellation than A1 without device. Similarly, at a CDF value of 0.5, configuration A2 with device is observed to achieve approximately 10 dB better cancellation than A2 without device. Hence, we conclude that placing antennas around a device improves the passive suppression by approximately 10 dB.

Second, we observe that at a CDF value of 0.5, configuration A2 with device achieves approximately 5 dB better cancellation than A1 with device. Similarly, A2 without device achieves approximately 5 dB better cancellation than A1 without device. Hence, we conclude that antenna placement A2 improves the passive suppression by approximately 5 dB with respect to antenna placement A1. The reason for this improvement is due to the fact that in A2 the receiver antenna main lobe is placed orthogonal with respect to the transmitter antennas main lobe. Consequently, A2 results in less coupling between self-interfering antennas and this results in larger levels of passive suppression.

Recent characterizations of passive suppression mechanisms ~\cite{Choi,Duarte} demonstrate levels of passive suppression lower than 60 dB. Our results in Figure~\ref{passive} show that taking into account the antenna pattern and placing the antennas around the full-duplex device serves as further means of passive suppression and helps achieve passive suppression values between 60 dB and 70 dB. Comparing the cancellation values for A1 without device and A2 with device in Figure~\ref{passive}, we observe that through device cancellation and orthogonal antenna placement improve the amount of passive suppression by approximately 15~dB.


\begin{result}[Impact of Passive Suppression on Self-interference Channel]
\emph{As the amount of passive suppression increases, the wireless self-interference channel becomes more frequency selective.}
\label{result:cancellation:2}
\end{result}

In our implementation of the analog canceler we compute the cancellation coefficient per subcarrier, $b_{i,m,n}[k]$, as shown in Eq.~(\ref{eqn:cancoeff}). Hence, $b_{i,m,n}[k]$ is the ratio of the estimate of the self-interference wireless channel $\widehat{h}_{i,m,n}[k]$ and the wire channel $\widehat{h}_{i,n}^{W}[k]$. Since the wire channel $\widehat{h}_{i,n}^{W}[k]$ is frequency flat, variations of the cancellation coefficient $b_{i,m,n}[k]$ as a function of the subcarrier index will be due to variations of the self-interference channel $\mathbf{h}_{i,m,n}$ as a function of frequency. If $\mathbf{h}_{i,m,n}$ is frequency flat then $b_{i,m,n}[k]$ will be the same across all subcarriers. If $\mathbf{h}_{i,m,n}$ is frequency selective then $b_{i,m,n}[k]$ will vary for different subcarriers.

Figure~\ref{cancoeff} shows the magnitude of the  cancellation coefficients, $b_{i,m,n}[k]$, for  each  of  the  48 data subcarriers captured for two subsequent packets. The subcarrier  spacing  is  0.3125  MHz  as in 802.11 for a  20  MHz  bandwidth channel.  We observe that as a function of subcarriers, the channel attenuation can vary significantly across frequency, and thus approximating self-interference channel as frequency flat can be highly inaccurate.

To completely characterize the statistical variations in self-interference channel across frequency, we use the measure of peak-to-peak (p2p) value of the magnitude of the cancellation coefficient, $|b_{i,m,n}|^{p2p}$, as follows,
\begin{equation}
|b_{i,m,n}|^{p2p}=\frac{\max_{k \in \{1,..., K\}}|b_{i,m,n}[k]|^2}{\min_{k \in \{1, ..., K\}}|b_{i,m,n}[k]|^2}.
\end{equation}
If the self-interference channel $\mathbf{h}_{i,m,n}$ is a flat frequency channel then $|b_{i,m,n}|^{p2p}=1$ and for a frequency selective channel $|b_{i,m,n}|^{p2p}$ will be larger than 1. For each FD2$\times$1 experiment we computed the value of $|b_{i,m,n}|^{p2p}$ between transmitter antenna 1 (T1) and receiver antenna 1 (R1). Figure~\ref{fig:CDF_cancoeff} shows a characterization of the cancellation coefficient for the four different antenna configurations listed in Table~I. Figure~\ref{fig:CDF_cancoeff} shows that the channel can have large variations in magnitude  in the practical case of antennas placed around a device, with a median of 9~dB p2p magnitude variations for A2 with device.

Comparing Figure~\ref{fig:CDF_cancoeff} with Figure \ref{passive}, we observe the following. The larger the passive suppression, the larger are the variations of the self-interference channel as a function of frequency. Intuitively this makes sense since passive suppression of the self-interference corresponds to suppression of the strongest line-of-sight paths between self-interfering antennas. As the line-of-sight path is weakened, the self-interference channel becomes more dependent on weaker reflected multi-paths and this results in larger frequency selectivity of the self-interference channel.

For scenarios where the channel is frequency-selective, the active analog cancellation must be able to adapt to the frequency variations of the channel per subcarrier, as is the case in our proposed implementation of active analog cancellation.





\subsection{Performance of Analog Cancellation}
\label{subsec:phy_canceller:analog}

To better illustrate the importance of the per subcarrier adaptation of the analog canceler, we compare the performance of our per subcarrier analog cancellation with the performance of two analog cancellation schemes that do not adapt the magnitude of the cancellation coefficient per subcarrier and use the same magnitude of the cancellation coefficient for all subcarriers (as is the case for the analog canceler schemes considered in \cite{Radunovic:2009aa,Jain,Choi}).

Specifically, we consider the following two flat-frequency cancelers -- (i) Flat-Frequency Canceller 1 (FFC1):  for this canceler the magnitude of the cancellation coefficient is the same for all subcarriers and is computed as the average from the required per subcarrier as $(1/K)\sum_{k=1}^{K}|b_{i,m,n[k]}|$, and (ii) Flat-Frequency Canceller 2 (FFC2): for this canceler the magnitude of the cancellation coefficient is the same for all subcarriers and is computed as the value required by the middle subcarrier in the band hence it is equal to $|b_{i,m,n}[K/2]|$. We highlight that the three analog cancelers, per-subcarrier, FFC1 and FFC2, are different only in the magnitude of the cancellation coefficient but have the same per subcarrier adaptation of the phase of the cancellation coefficient. The above simplification made our implementation easier for comparison while still allowing us to demonstrate the importance of per subcarrier adaptation.

Figure~\ref{fig:analog_canc} shows the amount of active analog cancellation that our proposed analog cancellation achieves for configurations A1 without device and  A2 with device and it also shows the performance of FFC1 and FFC2. We observe that per subcarrier adaptation of the magnitude of the cancellation coefficient achieves larger analog cancellation than FFC1 and FFC2. From the Figure~\ref{fig:analog_canc} we approximate that per subcarrier adaptation of the magnitude of the cancellation coefficient achieves approximately 5~dB larger cancellation than FFC1 and FFC2. Hence, we obtain the following result.

\begin{result}[Gains from Per-subcarrier Cancellation]
\emph{Per subcarrier analog cancellation improves the amount of analog cancellation, by approximately 5~dB, compared to cancelers which do not adjust the magnitude of the cancellation coefficient per subcarrier.}
\label{result:cancellation:4}
\end{result}

From Figure~\ref{fig:analog_canc} we observe that the analog cancellation was larger for the configuration without device compared to the configuration with device. Hence, the roles for best/worst cancellation are inverted with respect to what we had observed in Figure~\ref{passive}, where configurations with device showed better performance than configurations without device. The reason why the configuration with device achieves lower levels of analog cancellation is because analog cancellation is based on an estimate of the self-interference channel. The weaker the received self-interference (self-interference at the receiver antenna), the worse is the estimate of the self-interfering channel and the worse is the amount of analog cancellation achieved. Configurations with device have the weakest levels of received self-interference because they achieve the largest passive suppression. Hence, we have the following result.

\begin{result}[Passive impacts Analog]
\emph{As the amount of passive suppression increases, the amount of analog cancellation decreases.}
\label{result:cancellation:3}
\end{result}

The reasoning for Result~\ref{result:cancellation:3} was also noted in the simulation based analysis presented in~\cite{RiihonenDec2011} and the experiment based analysis of a narrowband canceller presented in \cite{Duarte:2011aa}. This paper extends our prior narrowband single-antenna result to wideband multiple antenna systems.

\subsection{Performance of Digital Cancellation}
\label{subsec:phy_canceller:digital}





We are now interested in characterizing the performance of digital cancellation. For this purpose, we quantify the amount of digital cancellation achieved when placing a digital canceller after each of the three analog cancelers analyzed in Figure~\ref{fig:analog_canc}. These results for digital cancellation are shown in Figure~\ref{fig:digital_canc}. We observe that, when digital cancellation is placed after analog cancellation, the amount of digital cancellation achieved after our proposed per subcarrier analog canceller is less than the amount of digital cancellation achieved after FFC1 and FFC2 cancelers. This behavior is due to the following result.

\begin{result}[Analog impacts Digital]
\emph{As the amount of analog cancellation increases, amount of digital cancellation decreases.}
\label{result:cancellation:5}
\end{result}


The reason for Result~\ref{result:cancellation:5} is that as the amount of analog cancellation increases, the residual self-interference decreases, hence there is more noise in the estimation of the residual self-interference after analog cancellation and this results in less digital cancellation. In the limit, if analog cancellation can achieve infinite dB of cancellation (perfect cancellation), then digital cancellation becomes unnecessary and applying digital cancellation in this limit case will only lead to an increase in the noise. The reasoning for Result~\ref{result:cancellation:5} was also noted in the simulation based analysis presented in~\cite{RiihonenDec2011} and the experiment based analysis of a narrowband single-antenna canceler presented in \cite{Duarte:2011aa}. More details on wideband multiple antenna experiment results that demonstrate Result~\ref{result:cancellation:5} can be found in~\cite{MelissaThesis}.

\subsection{Total Cancellation of Physical Layer Design}
\label{subsec:phy_canceller:total}


We now compare the performance of our physical layer design, which uses per subcarrier analog cancellation, with the performance of cancellation designs that do not use per subcarrier analog cancellation. Specifically, we compare the results for total cancellation for systems which have the same passive suppression and digital cancellation mechanisms but use different analog cancelers. The different analog cancelers being the ones analyzed in Figure~\ref{fig:analog_canc} (per subcarrier, FFC1, and FFC2). The results for total cancellation are show in Figure~\ref{fig:total_canc_PSvsFFC}. We observe that using per subcarrier analog cancellation achieves the largest total cancellation and the improvement is approximately 3~dB. We note that the advantage of our per subcarrier analog cancellation is, not only that it improves the total cancellation by 3~dB, but also that it achieves larger pre-ADC cancellation compared to the FFC1 and FFC2 systems. Predictably, we will show in Section~\ref{subsec:phy_rate:comparison}, larger per subcarrier analog cancellation results in larger rates than using FFC1 or FFC2 cancelers.

Next, we analyze the total cancellation of our  design for the four different antenna configurations showed in Table~\ref{tbl:antenna_config}. Figure~\ref{total} shows a characterization of the total cancellation achieved when combining passive suppression with active per subcarrier analog and digital cancellation. We observe that A2 with device achieves the largest total cancellation. The cancellation values for A2 with device are between 70 dB and 100 dB with a median of 85 dB. In general, we observe that for the same implementation of active analog and digital cancellation, the largest cancellation will be obtained with the configuration that achieves the largest passive suppression. This leads to an important direction that  \emph{antenna design and placement} are crucial for achieving practical full-duplex, and the design has to be cognizant of the device dimensions and placement. Finally, we observe that the performance of the cancellation scheme was very similar between the FD2$\times$1 and FD1$\times$1 systems.

To the best of our knowledge, the levels of cancellation achieved by our A2 with device implementation are the best reported for a wideband 20 MHz multiple subcarrier and multiple antenna full-duplex system. The results provided in~\cite{Choi,Duarte,Duarte:2011aa,Khojastepour,Aryafar} correspond to narrowband systems. The results in~\cite{Jain} are for a multiple subcarrier system with 10 MHz bandwidth and correspond to a single interference antenna. The work in~\cite{Khojastepour,Aryafar,Jain} does not report a measured value of total cancellation for a combination of passive, active analog, and active digital cancellation and focuses only on characterizing a subset of these types of cancellations. Finally, none of the previous works~\cite{Jain,Choi,Khojastepour,Aryafar,Duarte,Duarte:2011aa} report cancellations larger than 73~dB. Hence, we have the following result.
\begin{result}
\emph{Our proposed self-interference canceller design, for 20MHz  FD1$\times$1 and FD2$\times$1 systems,  achieves total self-interference cancellation values similar or larger than what prior work has reported. }
\label{result:cancellation:7}
\end{result}


Finally, we characterize the amount of residual self-interference to noise ratio. The residual self-interference is the amount of self-interference left after all the cancellation stages (passive, active analog, and active digital) have been applied. Figure~\ref{fig:residualinr} shows the residual self-interference to noise ratio (INR). As expected, configuration A2 with device results in the lowest levels of residual INR since this configuration is the one that achieves the largest cancellation. Although our self-interference canceller design can achieve larger cancellation than what related work has reported, we observe from Figure~\ref{fig:residualinr} that these cancellation values are not enough to guarantee that the self-interference is reduced to the noise floor (INR=0 is not guaranteed). However, as we will show in Section~\ref{subsec:phy_rate:comparison}, there are conditions under which full-duplex can achieve higher rates than half-duplex even if the self-interference is not reduced to the noise floor.


\section{PHY Evaluation: Rate Performance}
\label{sec:phy_rate}

\subsection{Metric for PHY Rate Analysis: Empirical Ergodic Rates}
\label{subsec:phy_rate:Metric}

The ergodic rate is the fundamental measure of PHY layer capacity in fading channels~\cite{Tse} and is an upper bound on the throughput that would be achieved by any MAC protocol. The ergodic rates become the starting point for a system designer to choose actual constellation sizes and code rates. The ergodic rate (ER) for transmission to Node $i$ is given by ${\mathbb E}\left( R_i\right) = {\mathbb E}\left[\log(1+ \mbox{SINR}_i[p])\right]$ where the expected value is computed as the average over all the packets $p$ transmitted to Node $i$ and $\mbox{SINR}_i[p]$ is the post processing Signal-to-self-Interference-plus-Noise-Ratio for packet $p$ received at Node $i$.

The empirical ergodic rate in experiments is computed based on an estimate of $\mbox{SINR}_i[p]$. We estimate $\mbox{SINR}_i[p]$ from transmitted and received constellation symbols as follows. The constellation symbol $s_i$ is sent to Node $i$ via the wireless channel. Node $i$ processes the received signal and computes $\widehat{s}_i$ which is the estimate of $s_i$. The average energy of the error or noise is given by ${\mathbb E} \left[|s_i-\widehat{s}_i|^2\right]$. Post processing SINR for packet $p$ received at node $i$, $\mbox{SINR}_i[p]$, is computed as $\mbox{SINR}_i[p] = \frac{{\mathbb E} \left[|s_i|^2\right]}{{\mathbb E}\left[|s_i - \widehat{s_i}|^2\right]}$ where the expected value is computed as the average over all the symbols transmitted to Node $i$ during packet $p$.

Since the two-way communication in half-duplex is achieved by time sharing the link with a fraction of time $\beta$ dedicated for transmission from Node 1 and a fraction of time $1-\beta$ dedicated for transmission from Node 2, the ergodic rate for each node in a half-duplex two-node communication system has to be scaled by their time of transmission, leading to ${\mathbb E}\left(R_1^{HD}\right) = \beta{\mathbb E}\left( R_1\right)$ and  ${\mathbb E}\left(R_2^{HD}\right) = (1-\beta){\mathbb E}\left(R_2\right)$. We performed only symmetric experiments where $\beta=0.5$. For full-duplex transmissions, since both nodes transmit at the same time, the ergodic rate for each node in a full-duplex communication system is given by ${\mathbb E}\left(R_1^{FD}\right) = {\mathbb E}\left(R_1\right)$ and  ${\mathbb E}\left(R_2^{FD}\right) = {\mathbb E}\left(R_2\right)$.

\subsection{Comparison of Full-duplex and Half-duplex Ergodic Rates}
\label{subsec:phy_rate:comparison}

Previous work on full-duplex implementation~\cite{Jain,Choi,Aryafar} had not considered the case of placing antennas around the full-duplex device. In Section~\ref{sec:phy_canceller}, we demonstrated that placing the interfering antennas around the full-duplex device can improve the total cancellation by 10~dB with respect to the case where antennas are not placed around a device. In this section, we show that the increase in total cancellation results in full-duplex rate gains at WiFi ranges.

Experiment results in Figure~\ref{fig:rates} show the ergodic rate for transmission to a node in a two-way link as a function of the average received signal to noise ratio at a node (SNR). Experiment results in Figure~\ref{fig:rates_withdevice} correspond to the case of antennas placed around the device (configurations A1 and A2 with device in Table~\ref{tbl:antenna_config}). Experiment results in Figure~\ref{fig:rates_nodevice} correspond to the case of antennas placed without the device in the middle (configurations A1 and A2 without device in Table~\ref{tbl:antenna_config}). For each FD/HD system and antenna configuration, we ran two experiments for Links 1, 6, 7, 8, 9, and 10, and we ran one experiment for Links 2, 3, 4, 5. Hence, for each FD/HD system and antenna configuration we ran 16 experiments (an experiment consisted in transmitting 90 packets from each of the two nodes in the link). Results in Figure~\ref{fig:rates} are per node per link per experiment per system. Hence, Figure~\ref{fig:rates_withdevice} shows 64 markers for each system, which corresponds to 2 antenna configurations (A1 without device and A2 without device) $\times$ 16 experiments $\times$ 2 nodes. Similarly, Figure~\ref{fig:rates_nodevice} shows 64 markers for each system.

For each system depicted in Figure~\ref{fig:rates}, we present experiment results with markers and also show a line which corresponds to a linear fit of the experiment results. The linear fit for the FD2$\times$1, FD1$\times$1, and HD2$\times$2 systems was computed based on the data points that lied between 5~dB and 30~dB SNR. We only used this interval for the fit because most of the data points lie inside this interval and because this interval contains SNR values typical of WiFi operation (SNR $\leq 30$~dB). For the HD2$\times$1 and HD3$\times$1 systems the linear fit was computed based on the data points that lied between 5~dB and 23~dB. The reason we used this interval for the HD2$\times$1 and HD3$\times$1 linear fits is because we observe that for SNR values above 23~dB the rate of HD2$\times$1 and HD3$\times$1 systems does not increase as the SNR increases. This leads us to conclude that our HD2$\times$1 and HD3$\times$1 implementations reach a performance ceiling at SNR~$\approx 23$~dB. We have observed that the performance ceiling of our implementation is a function of the bandwidth. When we use only one subcarrier (625~kHz bandwidth system) we do not observe a performance ceiling. We have not observed a performance ceiling for our FD or HD2$\times$2 implementations.

For the experiments results in Figure~\ref{fig:rates}, we computed the SNR as the ratio of the average signal of interest RSSI at the node to average noise power. The average RSSI at a node was computed by averaging the measurements of received signal of interest power per packet over the 90 packets received during an experiment run. The average noise power was estimated to be $-90$~dBm for all experiments. The above estimate of noise power was based on the radio data sheet and on node calibration performed before the experiments were started. The ergodic rate was computed using per packet measurements as was explained in Section~\ref{subsec:phy_rate:Metric}.



We now analyze the rate performance of full-duplex and half-duplex systems that do not use antenna placement around the device. The rate results for these systems are shown in Figure~\ref{fig:rates_nodevice}. From Figure~\ref{fig:rates_nodevice} we observe that for SNR~$\leq 30$~dB, the experiment data points for the full-duplex systems are mostly below the experiment data points for half-duplex systems. Consequently, since the linear fits only consider experiment data points for which the SNR was lower than 30~dB, the linear fit for the FD2$\times$1 and FD1$\times$1 systems lies below the linear fit for the half-duplex systems. We observe that the experiment data points in Figure~\ref{fig:rates_nodevice} show that FD2$\times$1 and FD1$\times$1 systems can have similar or large rates than half-duplex systems at SNR values larger than 30~dB. However, SNR values larger than 30~dB are not typical in WiFi systems. Consequently, from results in Figure~\ref{fig:rates_nodevice} we conclude that full-duplex gains at WiFi ranges cannot be achieved with self-interference cancellation schemes that have a median cancellation of 78~dB or less, as is the case for the full-duplex systems without device considered in our experiments.


%

We next analyze the performance of full-duplex and half-duplex systems that use antenna placement around the device. By comparing the experiment data and linear fits in Figure~\ref{fig:rates_withdevice} with the experiment data and linear fits in Figure~\ref{fig:rates_nodevice} we observe the following. Placing antennas around the device (device in the middle) improves the performance of full-duplex and half-duplex systems with respect to the case where antennas are not placed around the device. In other words, for the same SNR, full-duplex and half-duplex rates tend to be higher when the antennas are placed around the device. For the half-duplex systems this is explained by the fact that placing antennas around the device reduces the transmitter and receiver correlation and as these correlations decrease, the half-duplex rates increase~\cite{cond,Tse}. For the full-duplex systems this is explained by the fact that placing the antennas around the device achieves larger cancellation than the configuration without a device in the middle of the antennas.

The larger self-interference cancellation achieved by placing the antennas around the device yields the following result.

\begin{result}
\emph{FD1$\times$1 and FD2$\times$1 systems with antenna placement around the device can consistently achieve larger rates than HD2$\times$2, HD3$\times$1, and HD2$\times$1 systems for SNR range of 20--30dB, thereby covering nearly half the range of a typical WiFi system.}
\label{result:rate:2}
\end{result}

Result~\ref{result:rate:2} can be verified from Figure~\ref{fig:rates_withdevice}. The linear fit for FD2$\times$1 and the linear fit for HD2$\times$2 in Figure~\ref{fig:rates_withdevice} shows that FD2$\times$1 can achieve rates larger than HD2$\times$2 for  SNRs  approximately higher than $20$~dB. Similarly, the linear fit for FD1$\times$1 and the linear fit for HD3$\times$1 in Figure~\ref{fig:rates_withdevice} show that FD1$\times$1 can achieve larger rates than HD3$\times$1 for SNRs approximately higher than $20$~dB.

From Figure~\ref{fig:rates_withdevice} we note that, for the case where antennas are placed around a device, the full-duplex systems have a larger multiplexing gain than all half-duplex systems, where multiplexing gains are the slopes\footnote{If a system has multiplexing gain of $r$ then at high SNR the rate in bps/Hz can be approximated as $r \log_2 \mbox{SNR}$ and the slope of increase, in a plot where the y axis is in bps/Hz and x axis is in dB, is equal to $(r \log_2 \mbox{SNR})/(10*\log_{10} \mbox{SNR})=0.332r$} of the rate-SNR curves in Figure~\ref{fig:rates}. For the case of antennas without a device, from Figure~\ref{fig:rates_nodevice} we observe that most of the full-duplex rates lie below 1.5 bps/Hz which explains the lower slopes of FD systems compared to HD systems when the antennas are placed without a device in the middle and when we consider SNRs lower than 30~dB for the linear fit. Thus, we obtain the following result.

\begin{result}
\emph{FD1$\times$1 and FD2$\times$1 systems were measured to have a larger multiplexing gain per-node than HD2$\times$2, HD3$\times$1, and HD2$\times$1 systems for the case where the antennas are placed around the device (device in the middle).}
\label{result:rate:3}
\end{result}

In Figure~\ref{fig:rates_withdevice} the slope of the fit for the FD1$\times$1 system is approximately 1.3 times larger than the slope of the fit for the HD2$\times$2 system and approximately 3.7 times larger than the slope of the fit for the HD3$\times$1 and HD2$\times$1 systems. The slope of the fit for the FD2$\times$1 system is approximately 1.5 times larger than the slopes of the fit for the HD2$\times$2 system and approximately 4.2 times larger than the slopes of the HD3$\times$1 and HD2$\times$1 systems.

FD1$\times$1 and FD2$\times$1 systems have a larger slope than HD3$\times$1 and HD2$\times$1 systems because of the following reason. At high SNR, the HD3$\times$1 and HD2$\times$1 systems can have a maximum multiplexing gain per-node of 0.5, since each node only transmits half the time. In contrast, at high SNR the full-duplex systems can have a maximum multiplexing gain per-node of 1, which is not scaled by 0.5, since full-duplex systems can transmit during the entire time slot. Thus, we conclude that FD1$\times$1 and FD2$\times$1 systems are expected to have steeper slopes for rate increase vs.\ SNR compared to HD2$\times$1 and HD3$\times$1 systems. We think that a reason why the fit for the FD systems achieves more than twice the slope of HD3$\times$1 and HD2$\times$1 systems is due to the performance ceiling of our implementation of HD3$\times$1 and HD2$\times$1 which was explained earlier.

As evident from the results in Figure\ref{fig:rates_withdevice}, FD1$\times$1 and FD2$\times$1 rates can also have a larger slope than a HD2$\times$2 system, which is a little surprising but we conjecture the following reason. As discussed above, the full-duplex systems can have a maximum multiplexing gain of 1. At high SNR, a 2$\times$2 MIMO system can have a maximum system multiplexing gain of 2, which implies that HD2$\times$2 can have a maximum multiplexing gain per-node of 1 (since each node transmits only half the time). As shown in~\cite{cond}, for a 2$\times$2 system to achieve maximum multiplexing gain, the mean channel condition number should be approximately 3.5~dB or less (see Figure 6 in \cite{cond}, where values of $\rho_{T}=\rho_{R}=0$ lead to a multiplexing gain of 2 at high SNR and from equations (10) and (14) we compute that $\rho_{T}=\rho_{R}=0$ correspond to a condition number of 3.5~dB, computed as $10\log_{10}\sqrt{(4-0.65)/0.65}$).  
From both our results and the results in~\cite{cond}, it appears that indoor channels typically do not have condition numbers which meet the above criterion, therefore reducing achieved multiplexing gains of MIMO systems compared to theoretical maximum. Figure~\ref{fig:CNCDF} shows the CDF of the condition number measured per packed from our experiment measurements. From Figure~\ref{fig:CNCDF} we observe that the probability of having a condition number less than or equal to 3.5~dB is small (less than 10\%)

Next, we comment on the importance of per-subcarrier cancellation. The lower active analog cancellation achieved by FFC1 and FFC2 in a frequency selective environment (see Figure~\ref{fig:analog_canc}) results in a degradation of the achievable rate performance as shown in Figure~\ref{fig:AR_active_freq}. More specifically, we observe from results in Figure~\ref{fig:AR_active_freq} that at an SNR of $15$~dB, schemes FFC1 and FFC2 result in approximately 13\% performance loss compared to our proposed scheme.

Finally we note that, although through-device cancellation is a simple and effective means of cancellation, antenna placement techniques in~\cite{Choi,Khojastepour,Aryafar} may not be able to implement and take advantage of through-device cancellation. The reason is that antenna placement techniques in~\cite{Choi,Khojastepour,Aryafar} are designed under the assumption that the self-interference channel does not have multi-path components. As we showed in Section~\ref{subsec:phy_canceller:passive}, placing the antennas around the device makes the multi-path effect more severe. Consequently, the assumptions that should hold for antenna cancellations in ~\cite{Choi,Khojastepour,Aryafar} to be effective may not hold (or hold for fewer realistic conditions) when the antennas are placed around a device.

%% file: mac_v.tex
\section{MAC Design}
\label{sec:mac}
In this section, we describe the design of the proposed MAC. The driving goal for full-duplex MAC design is minimal changes to the current WiFi standard, to support both half- and full-duplex nodes, and in the process, accelerate its adoption. We limited our attention to support full-duplex communication only between two nodes, i.e, when a mobile node and the AP have a packet for each other.
With that in mind, we modified the standard half-duplex 802.11 Distributed Coordination Function (DCF) WiFi MAC~\cite{Wifi-dcf} with RTS/CTS to add a full-duplex mode. In the rest of the section, we describe the  modifications to legacy WiFi MAC to support the full-duplex mode, while the complete details are provided in Appendix~\ref{apdmac}.

We divide our discussion on the changes to legacy WiFi MAC in three parts: (a)~discovery and transmission of full-duplex packets, (b)~management of ACKs and (c)~behavior of overhearing nodes.



\noindent
{\bf Discovery and transmission of full-duplex packets}:
The first challenge is to get an opportunistic full-duplex data transmission between two nodes. This is achieved with the following changes from the existing standard 802.11 modules. We make use of the standard Request to Send (RTS) and Clear to Send (CTS) packets. The sender (also known as the primary node) signals intent to send a data packet using a 802.11 RTS packet. The node receiving the RTS (secondary node) then discovers the transmit node id, which is needed to start the full-duplex transfer to the sender. Since the secondary node knows that the primary sender will be transmitting, it finds opportunity to send a full-duplex data as follows.

  In the standard 802.11 protocol, the RTS receiver sends a CTS frame and listens to the incoming data. However, in full-duplex MAC, we would also like the secondary node to transmit to the primary immediately after sending the CTS frame (after the standard SIFS time) whenever data is available. Note that at the head of the queue in the secondary node transmit buffer, there may not be a packet intended for the primary sender. We address this by making the secondary node inspect its queue and select the very first packet intended for the primary sender. If necessary, the secondary node further updates the Network Allocation Vector (NAV) based on the original NAV it received during the RTS frame and the length of secondary packet.

 In the standard 802.11 protocol, if a node receives a packet while transmitting, the data being transmitted is considered to have collided and it tries to retransmit the data. However, in our design both transmission and reception can occur at the same time, and hence a full-duplex exchange should not to be interpreted as a collision. Full-duplex MAC therefore checks if the received packet is from the other node involved in RTS-CTS exchange and accepts the packet if the NAV duration has not expired.

{\bf Management of ACKs}:
The second challenge for the two nodes involved in full-duplex data transmission is to send and receive acknowledgements for the successful transmissions.  In legacy 802.11, after sending the data, a node expects an acknowledgement (ACK) frame. However in full-duplex, since data is sent from both nodes simultaneously, each node gets data before it gets an ACK. This is fixed by accepting one data packet in the NAV duration, and returning to the prior state where  it is still waiting for ACK. However, on reception of a second data transmission before reception of ACK would result in the node concluding that the data has collided.

After the full-duplex transmission, both the nodes involved in transmission are waiting for an ACK. In legacy 802.11, nodes cannot transmit while waiting for ACK until the ACK timeout duration has passed. This would lead to neither of the two nodes receiving ACK (since the other node will not transmit) thus leading to the ACK getting timed out at both the nodes. In order to avoid this ACK timeout, full-duplex MAC allows the nodes to send an ACK even while waiting for ACK from the other end. Further, after sending the ACK, the node returns to the state where it is waiting for the ACK packet.

For asymmetric data packet lengths, the node sending a smaller length packet has to wait a long time to receive the ACK because the other node with larger packet length has not finished sending data. Thus in full-duplex MAC design, the wait time for the reception of the ACK packet for both nodes involved in the transmission is re-adjusted to the end of the NAV duration if necessary.

{\bf Behavior of overhearing nodes}:
The third challenge is to ensure that the node pair involved in full-duplex communication does not get an unfair advantage of capturing the channel repeatedly. In the standard 802.11 model, all nodes use a longer wait time (Extended Interframe space, EIFS) rather than the standard wait time (DCF Interframe Space, DIFS) if they receive an erroneous packet. The extended wait time is used to allow some other recipient for the packet who received the data correctly to be able to send an ACK frame in time. In full-duplex systems, all nodes that are not involved in full-duplex transmission will detect an erroneous packet and thus wait for EIFS. However, the nodes involved in full-duplex transmission do not find the reception erroneous and thus start their back-off timer after waiting for DIFS. Since the DIFS wait time is smaller than EIFS wait time, there is higher likelihood for one of the full-duplex nodes to grab the channel again causing unfairness.

    We propose two different strategies to solve this problem. The first is optimized for a system where all nodes are full-duplex. In full-duplex MAC design, full-duplex nodes that receive a CTS packet successfully that is not intended for them, ignore any erroneous packets during the NAV duration specified in CTS. Thus the nodes will not subsequently detect an erroneous packet caused by collision of two data streams and therefore not switch to using the EIFS mode. Note that the above change does not defeat the purpose of EIFS since the NAV time has already been understood by the node which includes the time taken for ACK transmission. 
    
     However, note that in a system with both full-duplex and \emph{legacy} half-duplex stations, as shown in the timing diagram in Figure \ref{mac}, the half-duplex nodes wait for EIFS rather than DIFS after a full-duplex transmission between AP and a full-duplex station in the system, leading to unfairness in the uplink throughput. To avoid unfairness, we propose a second strategy where full-duplex nodes do not ignore erroneous receptions during NAV except for the two nodes involved in full-duplex transmission to wait for EIFS after every full-duplex message exchange. Further, the two nodes involved in full-duplex transmission use a waiting time of EIFS rather than DIFS after receiving ACK. The change in waiting time will lead to increased wait times when there are a lot of full-duplex nodes while not compromising the uplink throughput of half-duplex stations significantly. We will analyze both strategies in Section~\ref{legacyeval}.

%% file: macsim2.tex
\section{MAC Evaluation}
\label{macsim}

In this section, we evaluate the full-duplex MAC protocol discussed in Section~\ref{sec:mac} using a commercial software package, OPNET Modeler-Wireless. We started with the standard 802.11 codebase available in OPNET and made the necessary modifications to implement full-duplex MAC. Figure~\ref{mac} shows a typical full-duplex framing structure where the RTS frame is followed by a SIFS time and a CTS frame. After another SIFS time, data transmission occurs simultaneously, and is then followed by a SIFS and  an ACK. The following timings in $\mu$sec are shown from our evaluated 802.11a-based full-duplex MAC implementation. For 18Mbps, DIFS=34$\mu$s, RTS=36$\mu$s, SIFS=16$\mu$s, CTS=32$\mu$s, Data=704$\mu$s, ACK=32$\mu$s. Note that after a full-duplex data transmission, the ACKs from both the nodes involved in the full-duplex exchange are transmitted simultaneously. 

First, we study how goodput varies with different modulation formats and varying packet sizes for a system with one AP and one full-duplex node (STA). Next, we extend our understanding regarding goodput performance for a system with multiple full-duplex nodes. Further, we study a system with a mix of full-duplex  and half-duplex nodes, where the half-duplex nodes ignore collisions during the NAV duration set by a RTS/CTS; this is non-standard behavior for legacy half-duplex nodes but serves to highlight the coexistence dynamics. Finally, we study the system with a mix of full and half-duplex nodes, where the half-duplex nodes are legacy nodes which do not ignore collisions in NAV duration.

Table~\ref{mactable} provides the simulation parameters used.
In all cases, we used AP to STA (node) distance separation of 14 m.
The free space path loss formula~\cite{proakis} was used to compute the amount of path loss. To model bit error rate, we assume that the self-interference and thermal noise are Gaussian, and we use the standard Q-functions for uncoded modulation~\cite{proakis}.
The maximum packet size is 1500 bytes; there is no segmentation and each packet is sent as the payload of one full-duplex MAC frame.

\subsection{Goodput characterization for different packet sizes}

We evaluate full-duplex MAC design for QPSK (18 Mbps)~\cite{ieee80211}. The rate value, 18 Mbps for QPSK, refers to the radio transmission rate.
For a two-node system with one AP and one node STA, we compare a full-duplex system against a legacy half-duplex system with or without RTS/CTS. Total goodput is defined as the sum of the goodput from the AP to the STA and from the STA to the AP. For half-duplex QPSK-18 Mbps, the total MAC goodput for half-duplex system without RTS/CTS is 13.69 Mbps, total goodput for half-duplex system with RTS/CTS is 12.8 Mbps while for full-duplex the MAC goodput is 25.62 Mbps for packet size of 1500 bytes.
Thus, there is a 87\% performance gain for full-duplex over half-duplex system without RTS/CTS, including all MAC overheads and 100\% gain for full-duplex over half-duplex system with RTS/CTS.

\begin{result}[Gains for symmetric and asymmetric traffic]
\emph{(a) Full-duplex MAC design doubles the throughput as compared to a legacy WiFi half-duplex with RTS/CTS, and increases by 87\% as compared to legacy WiFi half-duplex system without RTS/CTS for symmetric traffic.}

\emph{ (b) Full-duplex goodput is {\em always} higher than half-duplex goodput, where the half-duplex system may or may not use RTS/CTS.}
\label{result:mac1}
\end{result}

We now study the goodput performance when the traffic flow is asymmetric. Given that data traffic is predominantly based on TCP,
an asymmetric TCP download-only traffic will have 1500 bytes from the AP to the STA (downlink) and 40 bytes of TCP ACK packets going from the STA to the AP (uplink). We model asymmetry by varying the size of the packets from the STA to the AP, while keeping the cancellation fixed at 85 dB.
We note that it is well recognized that the performance gain of full-duplex is influenced by traffic asymmetry~\cite{contraflow}.

Table~\ref{pktsize} shows that as we increase the uplink packet size from 40 bytes to 1500 bytes, full-duplex provides a goodput gain ranging from 1.3x (13.14 vs 10.07) to 2x (25.62 vs 12.8) as compared to half-duplex system using RTS/CTS. The full-duplex goodput is {\em always} higher than half-duplex goodput, where the half-duplex system may or may not use RTS/CTS. As the length of uplink data packet decreases, the uplink throughput decreases for a full-duplex system while the downlink throughput remains the same. The uplink throughput is a factor of the size of data from node to AP divided by size of data from AP to node less than downlink throughput. In other words, the throughput divided by size of data in each direction is constant. In half-duplex system, the time-resource is shared by the two nodes while in full-duplex system, the downlink transmission always happens and the uplink transmission can be viewed as a bonus.


We now compare the effect of varying self-interference cancellation in Figure \ref{newfig} assuming symmetric 1500 byte data in both directions. We note that as the self-interference cancellation improves, higher constellations can be supported for full-duplex system thus improving the rate. The rate for half-duplex system does not depend on the self-interference cancellation. In the rest of this paper, we will mainly illustrate the results on 18 Mbps QPSK constellation (1500 bytes packet size) noting that the improvement factors will be similar at a fixed self-interference cancellation value of say 82-83 dB.

\subsection{Goodput characterization for multiple full-duplex nodes}
In this section, we consider the effect of scaling the network size to indicate multiple nodes with the packet sizes for all nodes being 1500 bytes.

\begin{result}[Gains for full-duplex system with multiple nodes]
\emph{The sum throughput for a full-duplex system with multiple nodes increases by a factor of approximately two when compared to a half-duplex system with RTS/CTS handshake.}
\end{result}

Consider the case when there are $n$ full-duplex nodes and one AP. In a 802.11 system, each node has equal chance of winning the channel contention due to random backoff and thus each of the $n+1$ node accesses the channel with equal probability.

For comparison of full- and half-duplex throughput, we first ignore all timing and collision overheads and assume that all nodes have infinite queues that are always saturated and have data to be transmitted. Further, the destination of each new arriving packet at the AP is uniformly distributed across all nodes. In a half-duplex system, each node grabs the channel for a fraction $1/(n+1)$ time. When a node grabs the channel, it sends data to AP and thus, the fractional time-slots used for data transfer from node to AP is $1/(n+1)$. When AP grabs the channel, it sends the data at the head of the queue to the appropriate node. Since the head of the queue can be addressed to any node with equally probability, AP sends packet to a node for $1/(n(n+1))$ of the time. The total throughput is normalized such that the sum throughput for the legacy half-duplex system is 1.

For a full-duplex system, each node grabs the channel for a fraction $1/(n+1)$ time. At each time, there are two concurrent communication paths ongoing, and thus the normalized throughput is 2. The uplink throughput from any node and downlink throughput to any node is $1/n$ which means that each node transmits to AP for $1/n$ fraction of the time-slots and vice versa. Thus, we see that full-duplex improves the uplink throughput from a node from $1/(n+1)$ to $1/n$ which translates to an improvement factor of $1+1/n$. The downlink throughput to node on the other hand improves from $1/(n(n+1))$ to $1/n$, an improvement by a factor of $n+1$. Even though the total throughput improves by a factor of 2, for $n>1$, the downlink throughput to a node improves by a larger multiple as compared to the uplink throughput from a node. The analysis here is summarized in the rows corresponding to full- and half-duplex systems in Table \ref{tbl:anal}.

The above analysis ignored several aspects, e.g. collisions and different packet sizes. The half-duplex system without RTS/CTS has collisions of data frames while full-duplex system has collisions of RTS frames. The results from OPNET simulations provide realistic throughputs incorporating all overheads as shown in Table \ref{tbl:multiplenodemac}. We report the sum goodput for the system together with average uplink/downlink goodputs. We also report collision statistics, for a system with RTS/CTS, (e.g. percentage of RTS requests that collided as a percentage of total RTS transmissions). 
Percentage RTS collisions is specifically the ratio of RTS that were not successfully received divided by total RTS packets sent including retransmissions multiplied by 100. Similarly, for half-duplex system without RTS/CTS, the data suffers collision and thus needs to be retransmitted. We provide the percentage of collided data frames which is the percentage of data packets not successfully received.

The ratio of sum goodput for full-duplex to a half-duplex system should theoretically be a factor of 2. We enumerate the ratio of sum goodput for the following cases: 2 for $n=1$, 2.03 for $n=2$, 2.03 for $n=4$, 2.02 for $n=8$, when compared to half-duplex system with RTS/CTS. We note that the number of retransmissions for a full-duplex system are much lower compared to a half-duplex system with RTS/CTS, e.g.\  the number of RTS retransmissions reduce from 17.8\% in the half-duplex case for $n=2$ to 8.1\% for the full-duplex case. 

We further observe that half-duplex system with RTS/CTS achieves better throughput than without RTS/CTS for $n\ge 4$. RTS/CTS incurs timing overheads that reduce the system throughput. However without RTS/CTS, a significant fraction of the time is wasted for recovering from the collisions of data frames which degrade throughput more than the collisions of RTS frames. Thus when $n\ge 4$, the collisions are high, and thus the time saved by having RTS frame collisions compared to having data frame collisions far outweigh the overhead of RTS/CTS signaling. Consequently, we use a half-duplex system with RTS/CTS as the baseline reference case for the rest of this paper.

From the above discussion, the uplink throughput from a node is a factor $n$ higher than the downlink throughput to a node for a half-duplex system. We  can be observe in Table \ref{tbl:multiplenodemac} that factor $n$ difference  holds approximately for the half-duplex system. We now compare the downlink throughput in full-duplex and half-duplex systems and note that theoretically, the full-duplex throughput should increase by a factor of $n+1$. Based on the results, the gain of downlink throughput from in a full-duplex system compared to a half-duplex system with RTS/CTS is $2$, $1.52$, $1.28$ and $1.15$  for $n=1$, 2, 4 and 8 respectively as compared to the theoretical 2, 1.5, 1.25 and 1.13 respectively. For the uplink case, the throughputs of full-duplex should increase by a factor of $(n+1)/n$ as compared to a half-duplex system. Our experimental results show improvement factor of $2$, $3.08$, $4.85$ and $8$ for $n=1$, 2, 4 and 8 nodes respectively compared to the theoretical values of 2, 3, 5 and 9 respectively. Thus, we find that the simulation results match well with theory.

%

\subsection{Coexistence for full-duplex and {\bf modified} half-duplex nodes that ignore collision during NAV}\label{sec:mod_hd}


We examine the performance when legacy half-duplex and full-duplex nodes coexist within the same wireless LAN. Suppose that there are $n=2m$ nodes comprising of $m$ full-duplex and $m$ half-duplex nodes. We first assume the case where the half-duplex nodes use a modified MAC design. The modified version provides an insight into the dynamics of co-existence, and also touches on an unfairness issue that we will highlight in the next subsection. Note that the modification may or may not be pragmatic in real scenarios.

We assume that the {\bf modified} half-duplex nodes ignore collisions during the NAV duration. Each node grabs the channel for a fraction $1/(n+1)$ time. When a half-duplex node grabs the channel, it sends data to AP. When a full-duplex node grabs the channel, AP also sends packets to full-duplex node. As a result, packets in the queue at AP for full-duplex nodes get depleted faster than the half-duplex nodes and does not have packets for full-duplex node with probability 1 resulting in the AP always transmitting to half-duplex nodes. Thus, the AP sends data to half-duplex nodes for fraction $1/(m(n+1))$ time-slots. The half-duplex node sends data to AP for a fraction $1/(n+1)$ time-slots. Further, full-duplex node sends data to AP and AP sends data to full-duplex node for a fraction $1/(n+1)$ time-slots.

Thus, the downlink throughput to half-duplex node increases from $1/(n(n+1))$ in a legacy half-duplex system to $1/(m(n+1))$ which is a factor of two improvement. The downlink throughput to full-duplex node increases from $1/(n(n+1))$ in a legacy half-duplex system to $1/(n+1)$ which is a factor $n$ improvement. Further, the uplink throughput from half-duplex/full-duplex node remains the same as a legacy half-duplex system. The overall sum throughput is thus $1+m/(n+1)$ factor more than the throughput of a half-duplex system. These results are summarized as Case 1 in the coexistence case in Table \ref{tbl:anal}.

\begin{result}[Coexistence with modified half-duplex nodes]
\emph{In a mix of $m$ full- and $m$ half-duplex nodes, the total throughput increases compared to a half-duplex-only system by a factor of $1+m/(2m+1)$}
\end{result}

In Table \ref{tbl:coexist_noeifs}, we see the goodputs for the case when there are $m$ full-duplex and $m$ half-duplex nodes. Case 1 represents the case where the modified half-duplex nodes avoid collisions during the NAV window. Note that the total throughput increases by a factor of $1.39$, $1.42$, and $1.45$ for $m=1, 2, 3$ respectively from the complete half-duplex system, as compared to $1.33$, $1.4$ and $1.43$ as predicted from the theory above.

Further, we note that the uplink throughput from any node to AP remains almost the same as the throughput from node to AP in a half-duplex node, which matches the above theoretical results. The downlink goodput to full-duplex node is the same as the uplink goodput from full-duplex node, which is also the same as the uplink goodput from a node in a legacy half-duplex system. Finally, the downlink throughput to half-duplex node should increase by a factor of 2 from the complete half-duplex system. We note that throughput improvement factor from the simulations is $1.8$ for $m=1$, $2.05$ for $m=2$, and $2.5$ for $m=4$, comparable to the theoretical results.

\subsection{Coexistence for full-duplex and legacy half-duplex nodes}\label{legacyeval}
In this subsection, consider legacy half-duplex nodes which will consider the full-duplex exchange as a collision and therefore would wait for EIFS rather than DIFS after every full-duplex transmission. The results for this scenario can be seen in Table \ref{tbl:coexist_noeifs}, Case 2. When the half-duplex nodes are legacy, we note that the full-duplex nodes wait for DIFS while the half-duplex nodes wait for EIFS after the end of every full-duplex transmission. The full-duplex nodes thus obtain an unfair advantage in accessing the channel. Thus, the throughput from the half-duplex node to the AP is reduced. Further, packets for full-duplex node are removed from the queue at the AP faster, and when AP successfully gets the channel, it has packets remaining only for half-duplex node. Therefore, the downlink to half-duplex and the goodput from AP to full-duplex node increase when the half-duplex nodes are legacy. Since full-duplex transmissions happen for a larger time-fraction, the overall sum goodput of the system increases. The above results were generated when the buffer at the AP is large. When the buffer at AP is limited, AP may not have packets for full-duplex node further limiting the throughput from AP to full-duplex node and thus the overall throughput reduces matching the performance closer to a half-duplex system.

For the scenario with legacy  half-duplex nodes, we propose a  coexistence based MAC change for full-duplex nodes such that the uplink throughput of half-duplex node is not greatly impacted. The change for the full-duplex nodes not involved in the full-duplex exchange is to wait for EIFS for full-duplex exchange like the legacy half-duplex nodes would be doing. The change for the full-duplex nodes involved in the full-duplex exchange is to wait for EIFS after every full-duplex data exchange like the legacy half-duplex nodes would be doing. The above change ensures that all nodes wait for DIFS before backoff after a half-duplex packet exchange, but wait for EIFS before backoff after a full-duplex packet exchange. As a result, the full-duplex nodes are more polite and do not get unfair advantage of repeatedly grabbing the channel. However, increased wait times after each full-duplex transmission is slightly disadvantageous when there are a lot of full-duplex nodes compared to legacy half-duplex nodes since the full-duplex nodes will wait longer (EIFS instead of DIFS). We note that the decrease in throughput due to extra wait time is not significant. The overall throughput for all full-duplex node decreases from $25.6$ Mbps to $25.05$ Mbps for $n=8$ and the throughput for $4$ full-duplex and $4$ half-duplex system reduce from the originally defined full-duplex MAC and slightly modified half-duplex MAC above from $18.37$ Mbps to $17.12$ Mbps for the ``graceful" full-duplex system with legacy half-duplex system. Even though the overall throughput decreases, the extra wait time based system does not compromise the legacy half-duplex nodes' uplink throughput significantly. Thus, we can summarize our main result of this section as follows.

\begin{result}[Coexistence of Modified full-duplex and legacy half-duplex nodes]
\emph{Overall system throughput decreases with extra wait time for full-duplex nodes, as compared to the coexistence between full-duplex and modified half-duplex nodes, but ensures minimal loss to legacy nodes' throughput.}
\end{result}


\subsection{ Extensions}

The full-duplex MAC can be extended to incorporate the following features.

{\em Beyond DCF:} 802.11 modes, such as HCF and frame aggregation, may be useful to aggregate smaller packets to achieve increased symmetry between the two directions of full-duplex.

{\em Full-duplex Data Available:} A CTS sender can indicate that it intends to send full-duplex data by re-purposing an unused bit (e.g. the More Data bit in the CTS Frame Control field \cite{ieee80211} section 7.2.1). Another option is for a full-duplex node to use two MAC addresses, one for half-duplex use, and the other for full-duplex use. This allows the RTS sender to adapt its rate accordingly. The new duration can be set as a NAV duration by the CTS sender.

{\em Three-node full-duplex:} The AP can receive from one node and transmit to another node as in \cite{Evan}. The second node needs to get a sufficiently stronger signal from the AP. If the AP knows the path losses between nodes, it can select a suitable node pairs, and power control can be used for both transmissions.

{\em No Tx/Rx switching:} The duration, e.g. between DATA and ACK, could potentially be reduced since there is no time needed to switch from receive to transmit.

{\em Intelligent choice of full-duplex vs.\ half-duplex:} The RTS and CTS packets offer an opportunity for the nodes A and B to estimate the signal quality which can allow for an informed choice of rate tuple and the full-duplex vs.\ half-duplex decision.

%% file: concl.tex

\section{Conclusions}
\label{sec:concl}
This paper presents the first design for a full duplex multi-antenna 20~MHz WiFi-ready design.
We achieved the best self-interference cancellation reported till date, and presented an integrated PHY and MAC design that is compatible with IEEE 802.11x, enabling accelerated adoption of FD wireless.
Our design achieves high rate and extended range, adequate for  most indoor WiFi deployments.
We believe that our results conclusively show that full-duplex WiFi is possible and can be highly beneficial in
practical propagation environments.

%% file: apdmac.tex
\appendix

\section {Sequence of steps for full-duplex MAC}\label{apdmac}

In this section, we describe a detailed sequence of steps for a full-duplex data exchange between an AP and a STA as per our design. The corresponding states (denoted by using \{\}) are shown in the flow chart in Figure~\ref{fdmac}.

\noindent {\bf 1)} An idle STA \{T1\} gets a packet from its higher layer which it needs to send to AP \{T1\}.

\noindent {\bf 2)} STA may need to wait \{T2\} if it has not waited already. STA listens on the desired channel.

\noindent {\bf 3)} If channel is idle (i.e. no active transmitters), STA sends a RTS frame \{T3\} to AP with the duration field specified in the network allocation vector (NAV) based on the frame length. After sending RTS \{T4\}, STA waits for a CTS response \{T5\}

\noindent {\bf 4)} Assuming correct transmission, AP goes from \{T1\} to \{R1\} to \{R2\} to \{R3\} while receiving the RTS. The AP gets the RTS \{R4\} and needs to send a CTS. If it has a packet to send to STA, full-duplex mode will be used. The scheduler in AP searches for a packet destined for STA and ensures that it is at head of the transmit queue for transmission \{T15\}. It prepares the CTS \{T16\} and waits for a Short Inter Frame Space (SIFS) period \{T2\}, and then sends a CTS frame \{T2\}.

\noindent {\bf 5)}  Any third node that overhears the CTS remains silent and ignores all collisions on the channel till the end of NAV duration in the CTS.

\noindent {\bf 6)}  STA starts receiving the CTS, goes through \{T5-R1-R2-R3-R7\} to get the CTS, and then prepares to send the data frame \{T6\}, waits for SIFS \{T3\} and then sends the primary FD data frame FDDATA1 \{T3\}.

\noindent {\bf 7)}  If the AP has data to send, after sending CTS \{T17\}, it prepares the full-duplex data frame \{T18\}, waits for SIFS \{T2\} and then transmits the secondary FD data frame FDDATA2 \{T3\}.

\noindent {\bf 8)} Both data frames FDDATA1 and FDDATA2 travel in different directions at the same time. Both nodes wait for ACK after sending data \{T8\}. Both nodes receive the data \{R5\} and need to respond with an ACK. Since they were expecting an ACK, they set an FDACK flag to remember to come back and wait for the ACK. Then they prepare the ACK \{T13\}, wait for SIFS time \{T2\} after any data transmission has finished, and send the ACK \{T3-T14\} and come back to wait for their expected ACK \{T8\}. The ACK timeout is adjusted to the end of the NAV to allow the other node to finish transmitting a data frame that may be longer.

\noindent {\bf 9)} At each node, when the ACK is received as expected \{R1-R2-R3-R6\}, the node waits for DIFS and for a random backoff period \{T2\} before contending for the channel for the next transmission.

%% file: figtab.tex
\cleardoublepage

\begin{figure*}[h]
\centering
\includegraphics[trim = 11mm 11mm 11mm 1mm, height=1.9in]{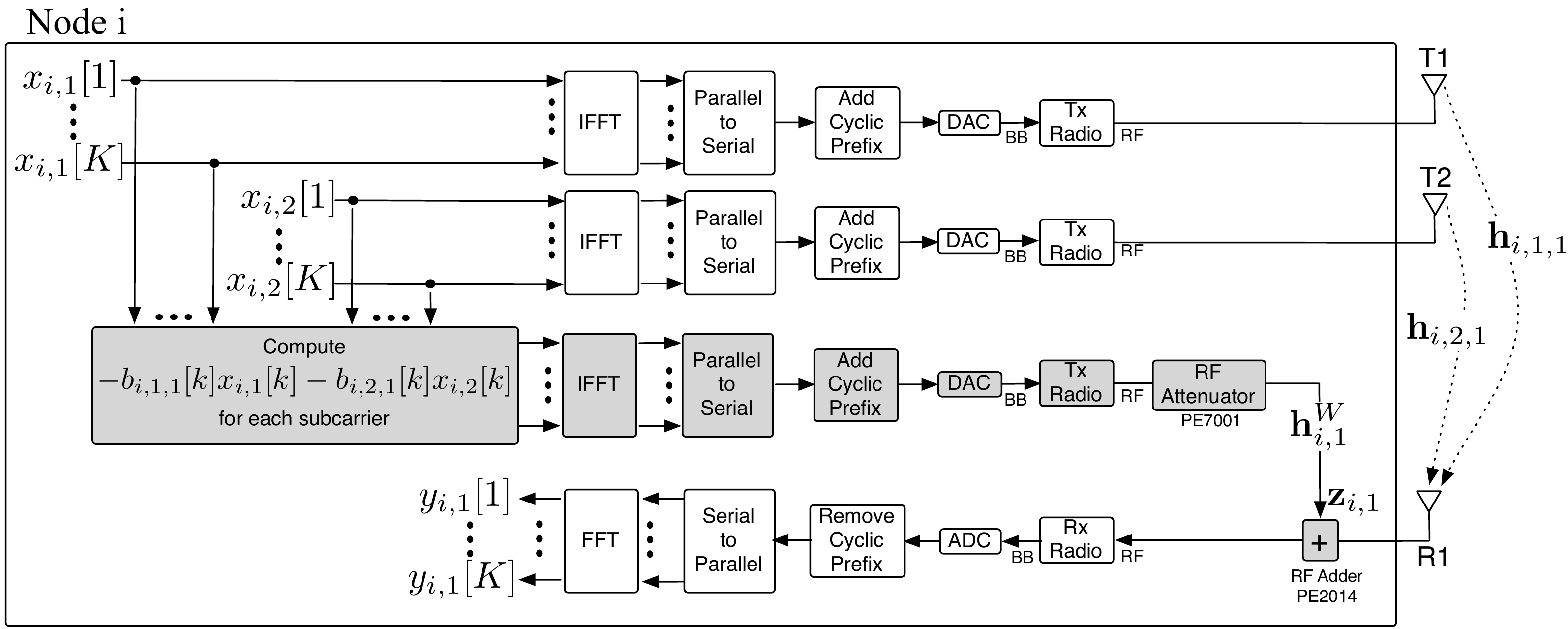}
\caption{Block diagram of a full-duplex OFDM node with two transmitter antennas and one receiver antenna (2$\times$1) using passive suppression and active analog self-interference cancellation. Blocks used for active analog cancellation are highlighted in gray. Passive suppression consists in propagation loss through $\mathbf{h}_{i,1,1}$ and $\mathbf{h}_{i,2,1}$. The Tx Radios are radio chains that up-convert from baseband (BB) to RF. The Rx radios are radio chains that down-convert from RF to BB.}
   \label{fig:fdnode}
\end{figure*}

\begin{table}[h]
\centering
\begin{tabular}{c|c|c|}
\cline{2-3}
& Antenna Placement 1  (A1) & Antenna Placement 2  (A2) \\
\cline{2-3}
& T2  is orthogonal  to T1/R2 & T2 is parallel  to T1/R2 \\
& R1/T3 is  parallel  to T1  & R1/T3 is  orthogonal  to T1 \\
\hline
\multicolumn{1}{|c|}{\rotatebox{90}{With Device}} & \includegraphics[scale=0.4]{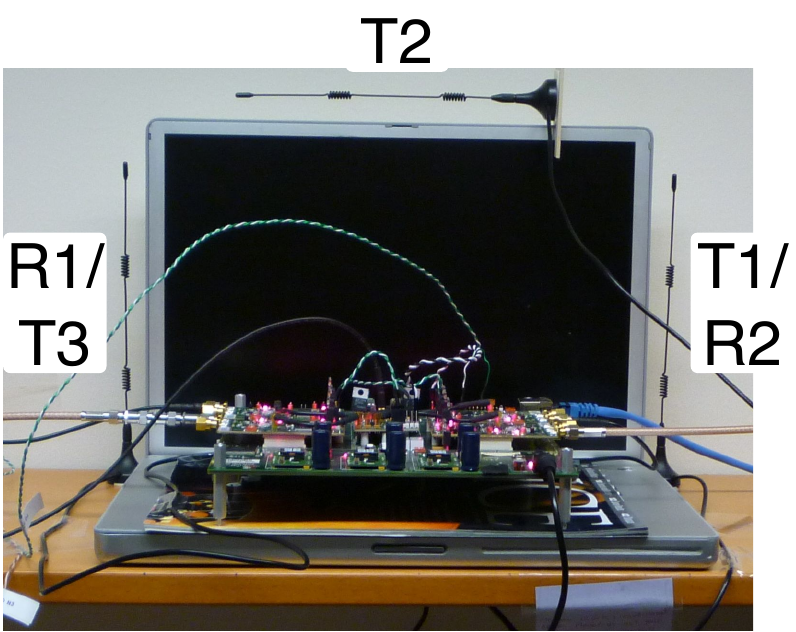}  & \includegraphics[scale=0.4]{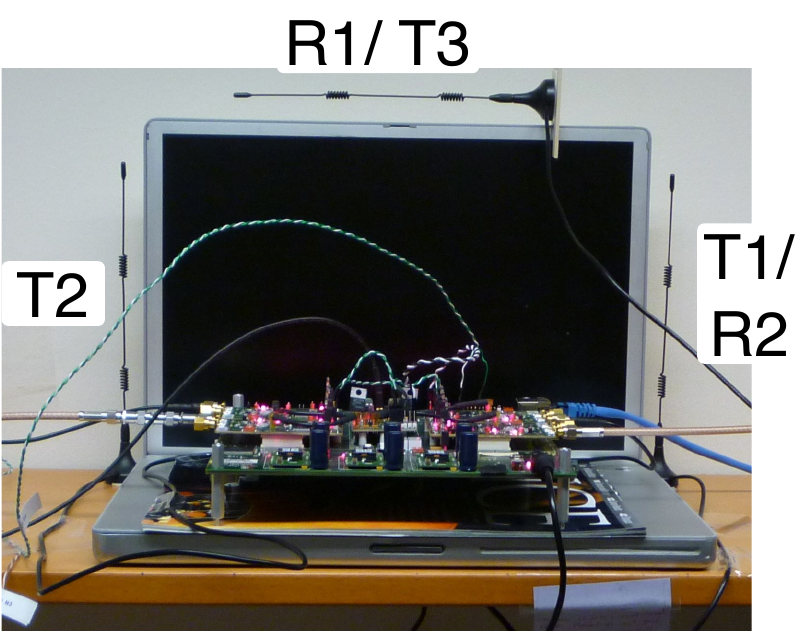} \\
\hline
\multicolumn{1}{|c|}{\rotatebox{90}{No Device}}  & \includegraphics[scale=0.4]{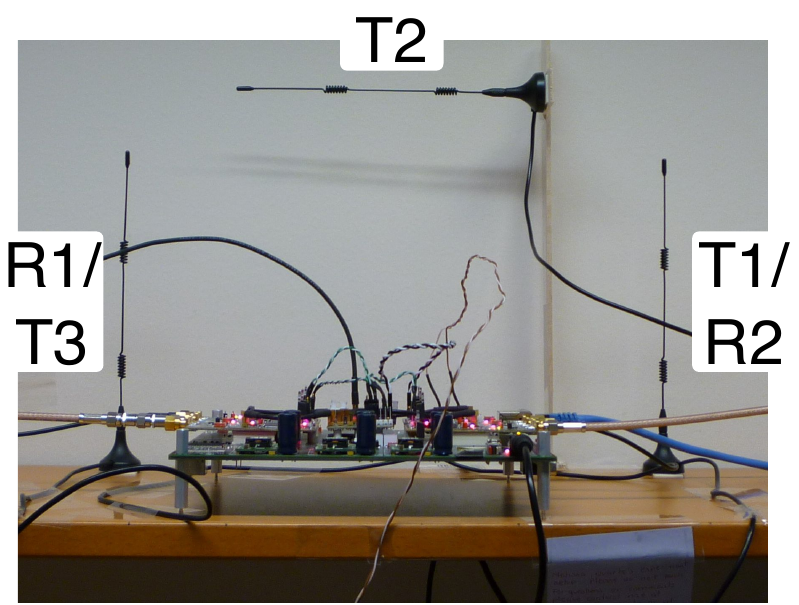}  & \includegraphics[scale=0.4]{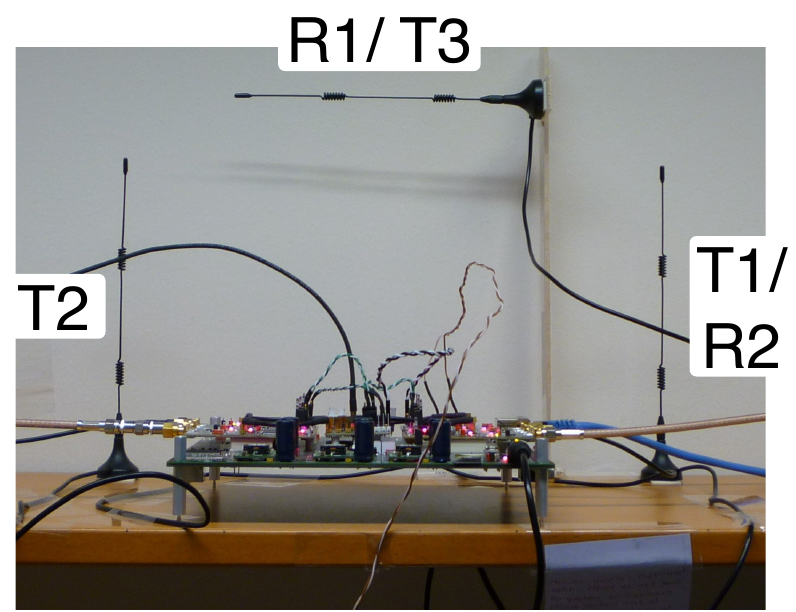} \\
\hline
\end{tabular}
\caption{Different antenna configurations with distance between parallel antennas of 37.5 cm}
\label{tbl:antenna_config}
\end{table}


\begin{figure}[h] 
   \centering
   \includegraphics[width=2.5in]{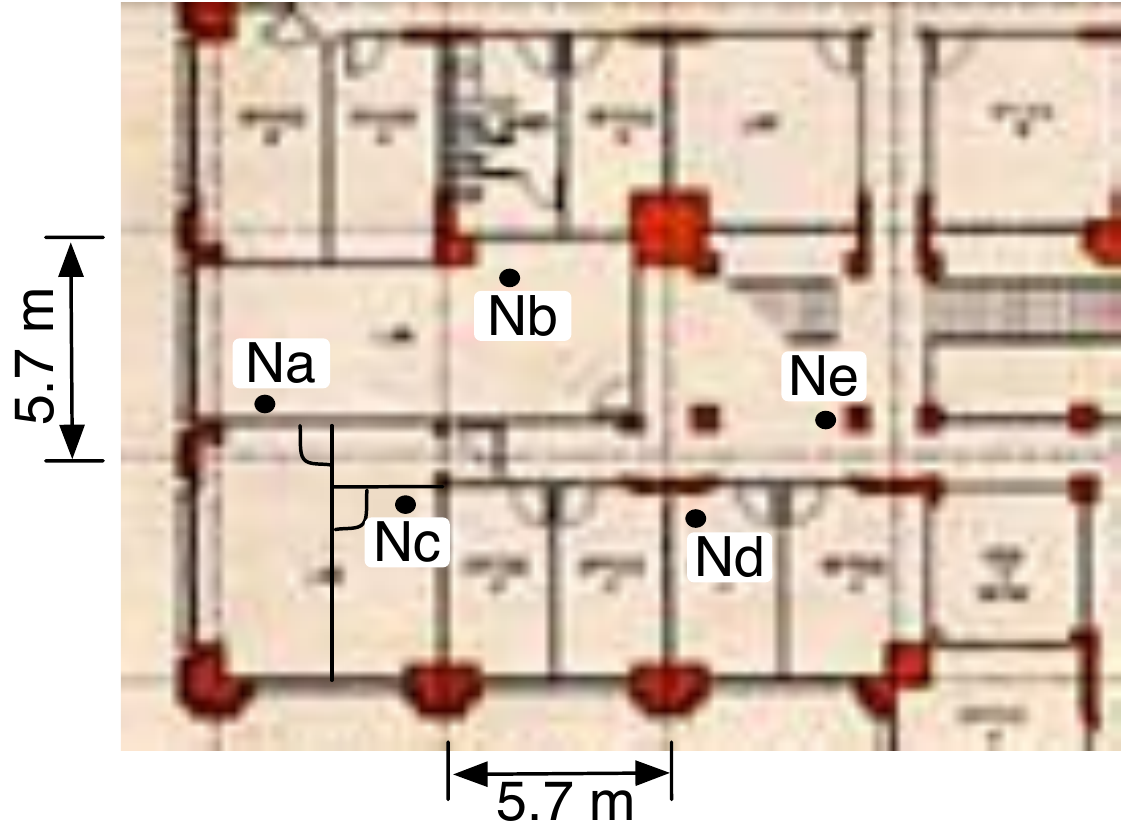}
   \vspace{-0.25in}
   \caption{Setup locations of nodes.}
  \label{setup_loc}
   \vspace{-0.1in}
\end{figure}

\begin{table}[h]
\centering
\begin{tabular}{ccccc}
\hline
Link & Node & Physical  & Type of  & Walls \\
number & pair & Distance (m) & channel & crossed \\
\hline
1 & Na, Nb & 6.5 & LOS & 0\\\hline
2 & Nd, Ne & 4.4 & NLOS & 1 \\
3 & Nb, Ne & 9.3 & NLOS & 1 \\
4 & Nc, Ne & 11.3 & NLOS & 1 \\
5 & Na, Ne & 14.8 & NLOS & 1 \\ \hline
6 & Nb, Nc & 6.5 & NLOS &  2 \\
7 & Nb, Nd & 8.3 & NLOS &  2 \\ \hline
8 & Na, Nc & 4.7 & NLOS & 3 \\
9 & Nc, Nd & 8.2 & NLOS & 3 \\
10 & Na, Nd & 12.7 &NLOS & 3 \\
\hline
\end{tabular}
\caption{Links considered in experiments}
\label{tbl:linkdistance}
\vspace{-0.1in}
\end{table}

\begin{table}[h]
\centering
\begin{tabular}{ccccc}
\hline
       & Number of & Number of          & Number of        & Total number\\
System & antennas  & up-converting radio & down-converting radio  & of radio chains\\
       & per node  & chains per node           & chains per node         & per node\\
\hline
FD 2$\times$1 & 3 & 3 & 1 & 4\\
HD 3$\times$1 & 3 & 3 & 1 & 4\\
HD 2$\times$2 & 2 & 2 & 2 & 4\\
\hline
FD 1$\times$1 & 2 & 2 & 1 & 3\\
HD 2$\times$1 & 2 & 2 & 1 & 3\\
\hline
\end{tabular}
\caption{Number of antennas and radios per node used by the different full-duplex and half-duplex systems that were evaluated via experiments.}
\label{tbl:numradios}
\end{table}

%

\begin{figure*}[h]
\centering
\subfigure[Passive Cancellation for four different antenna placements]{
\includegraphics[width=0.4\textwidth]{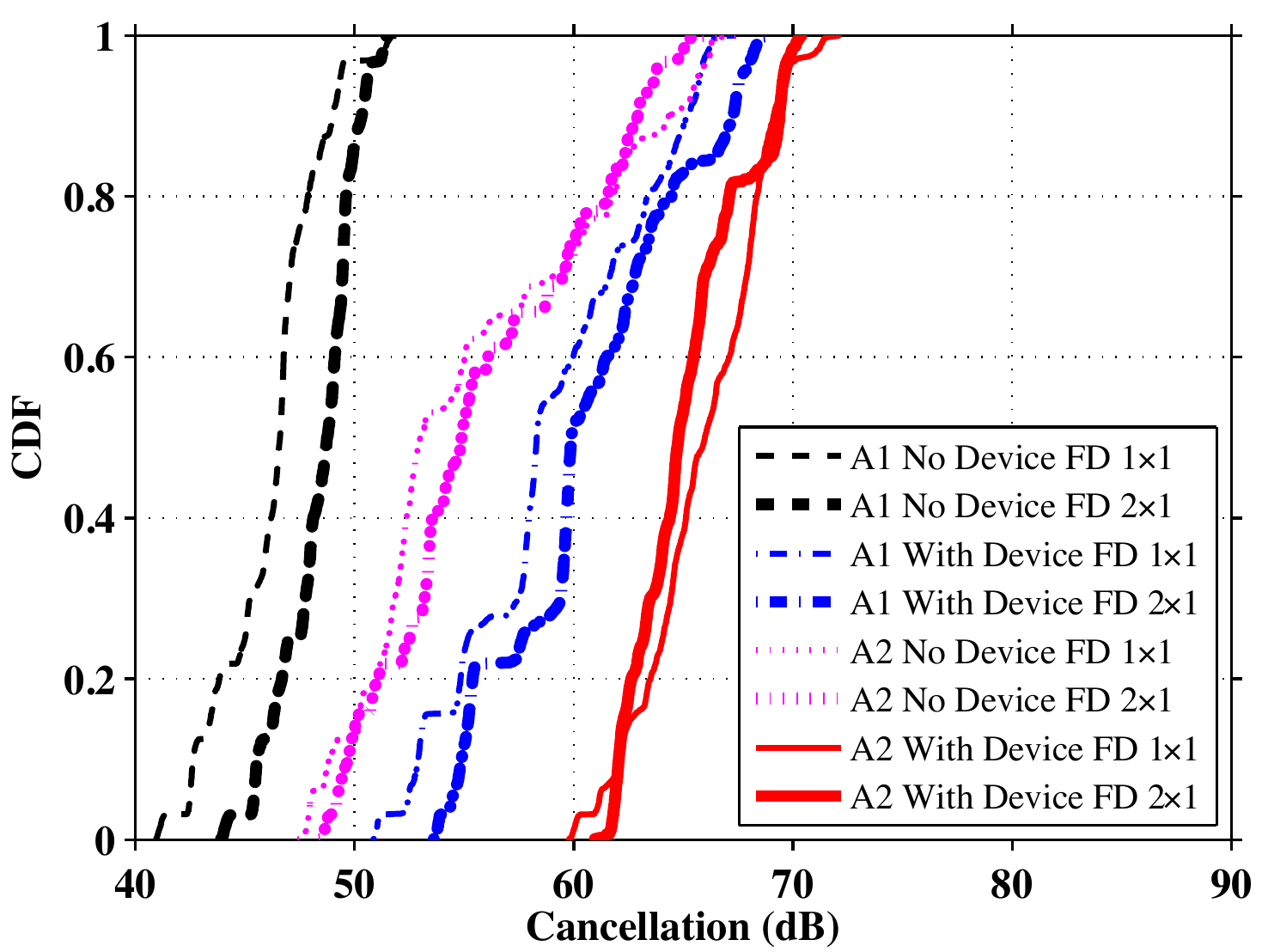}
   \label{passive}
}
\subfigure[Analog cancellation for different analog cancelers. The results are shown for two different placements.]{
\includegraphics[width=0.4\textwidth]{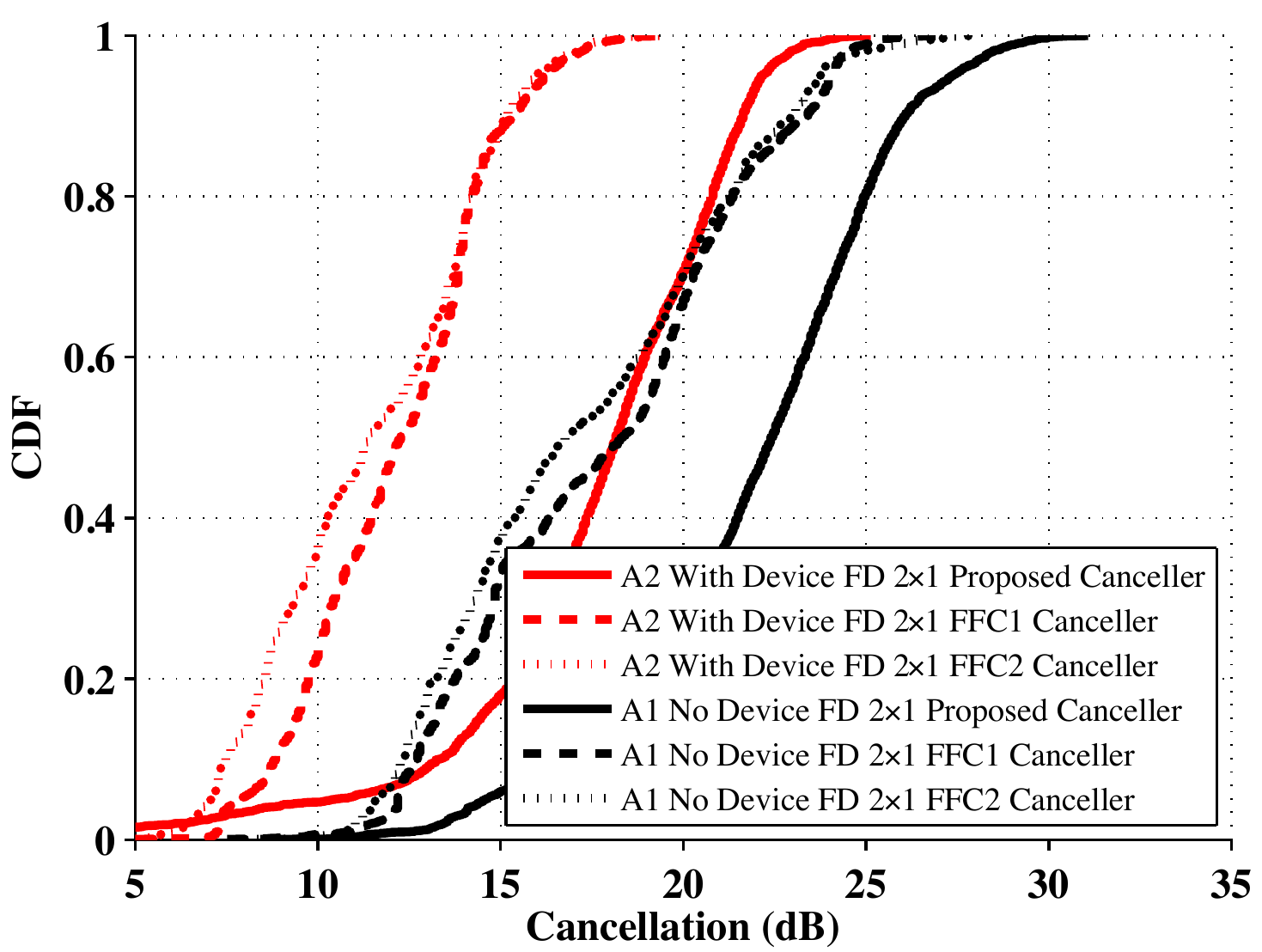}
   \label{fig:analog_canc}
}
\subfigure[Digital Cancellation when the same digital canceler is applied after different analog cancelers. The results are shown for two different placements.]{
\includegraphics[width=0.4\textwidth]{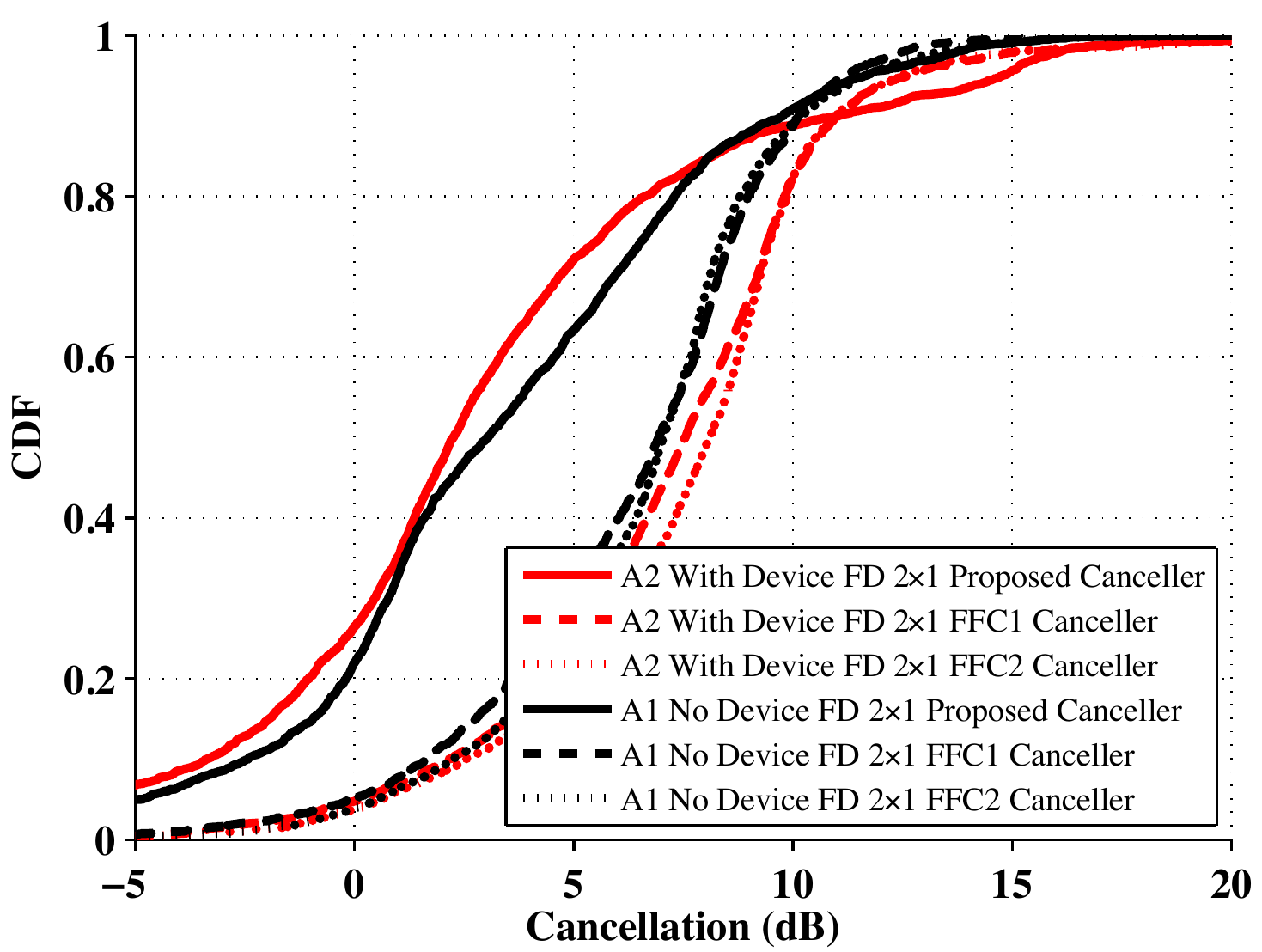}
   \label{fig:digital_canc}
}
\subfigure[Total cancellation for systems which have the same passive suppression and digital cancellation mechanisms but different analog cancelers. The results are shown for two different placements.]{
 \includegraphics[width=0.43\textwidth]{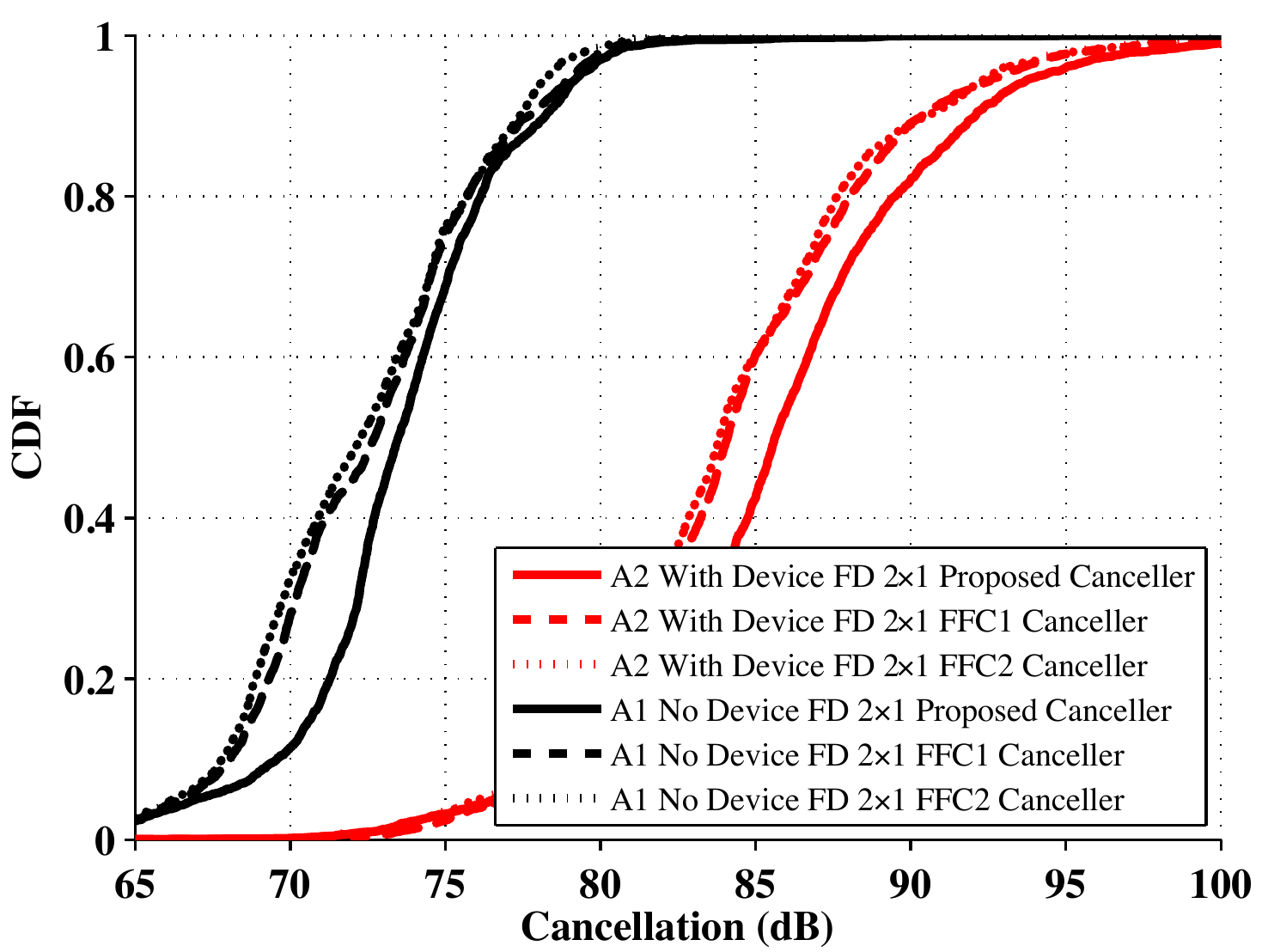}
   \label{fig:total_canc_PSvsFFC}
}
\subfigure[Total Cancellation for different antenna placements and using our proposed per subcarrier analog canceler and digital canceler.]{
 \includegraphics[width=0.4\textwidth]{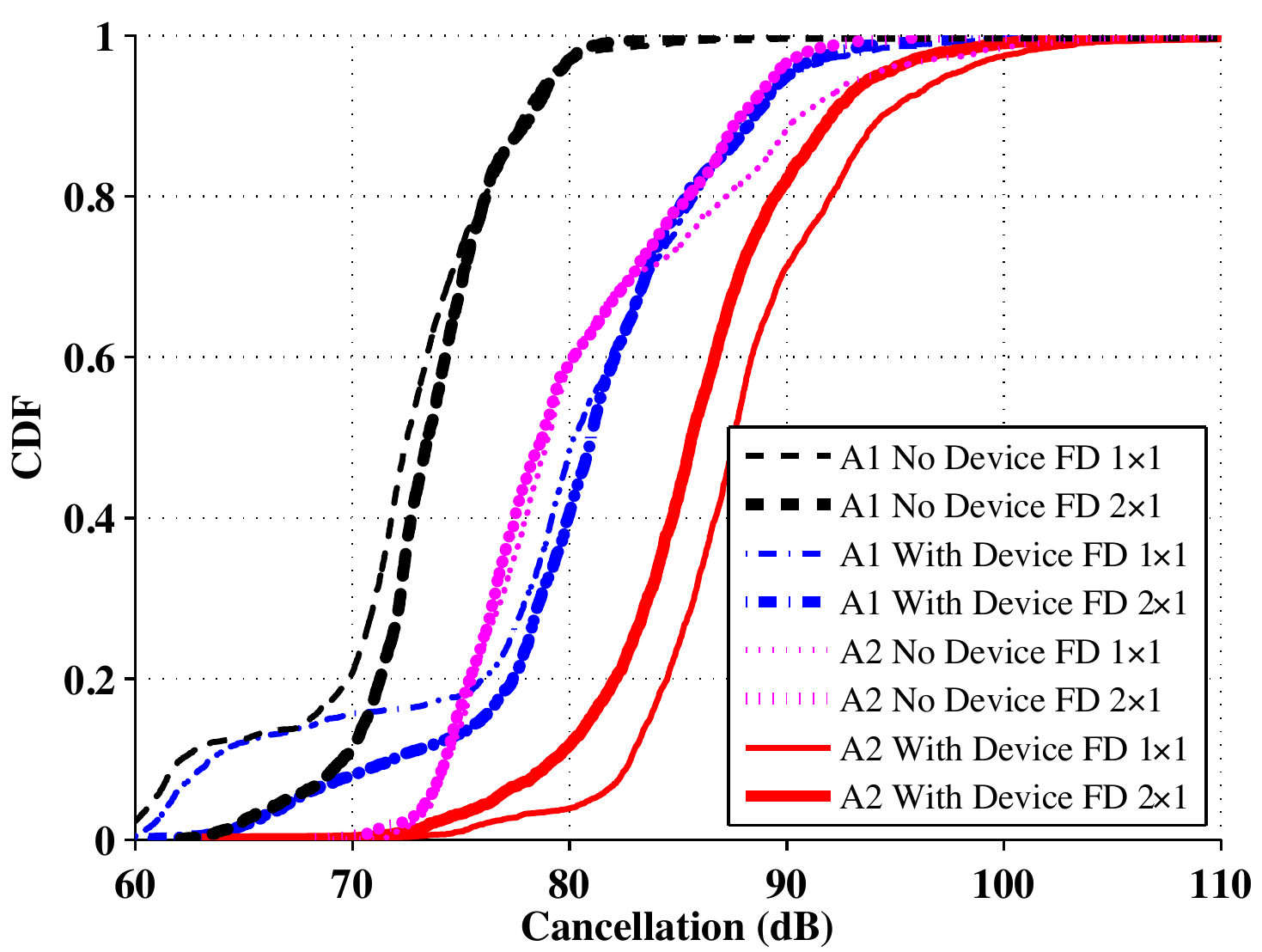}
   \label{total}
}
\caption[]{CDF of the amount of cancellation for different cancelers and different antenna placements.}
\end{figure*}


\begin{figure*}[h]
\centering
\subfigure[Cancellation coefficient per subcarrier captured for two subsequent packets.]{
\includegraphics[width=0.4\textwidth]{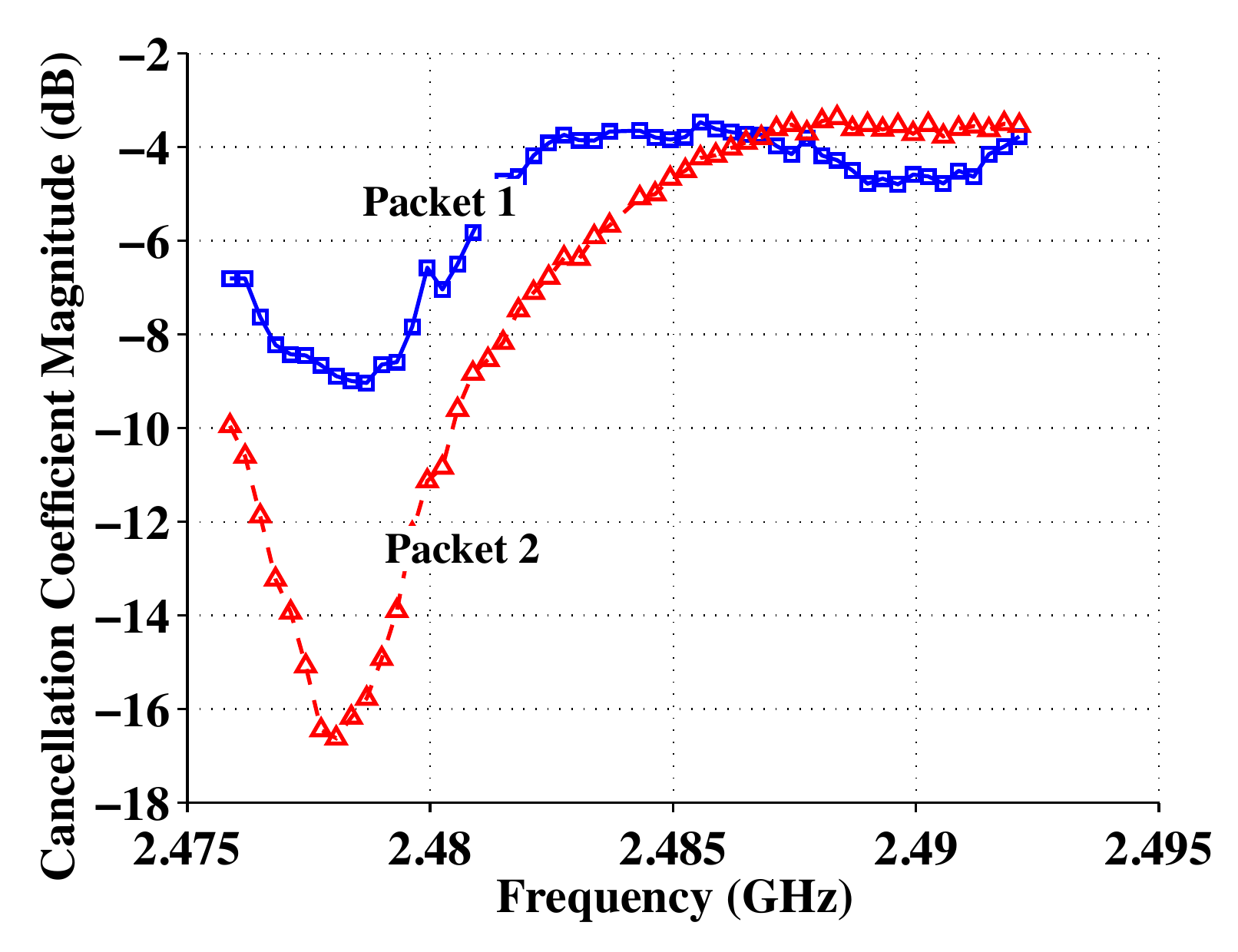}
   \label{cancoeff}
}
\subfigure[CDF of the peak-to-peak value of the cancellation coefficient magnitude.]{
 \includegraphics[width=0.4\textwidth]{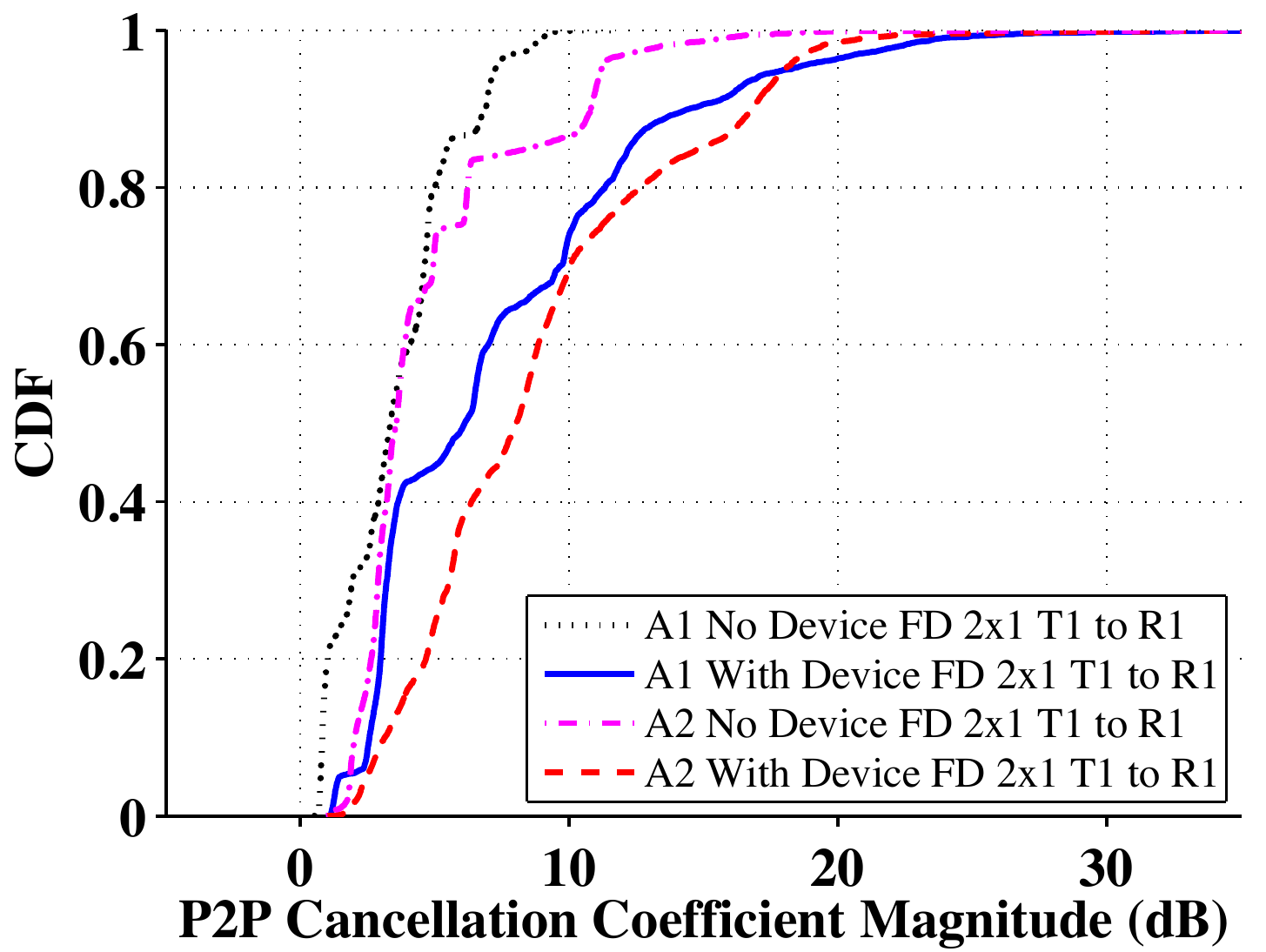}
   \label{fig:CDF_cancoeff}
}
\caption[]{Characterization of the effect of passive cancellation on the frequency response of the self-interference channel.}
\end{figure*}

\begin{figure}[h] 
   \centering
   \includegraphics[width=0.48\textwidth]{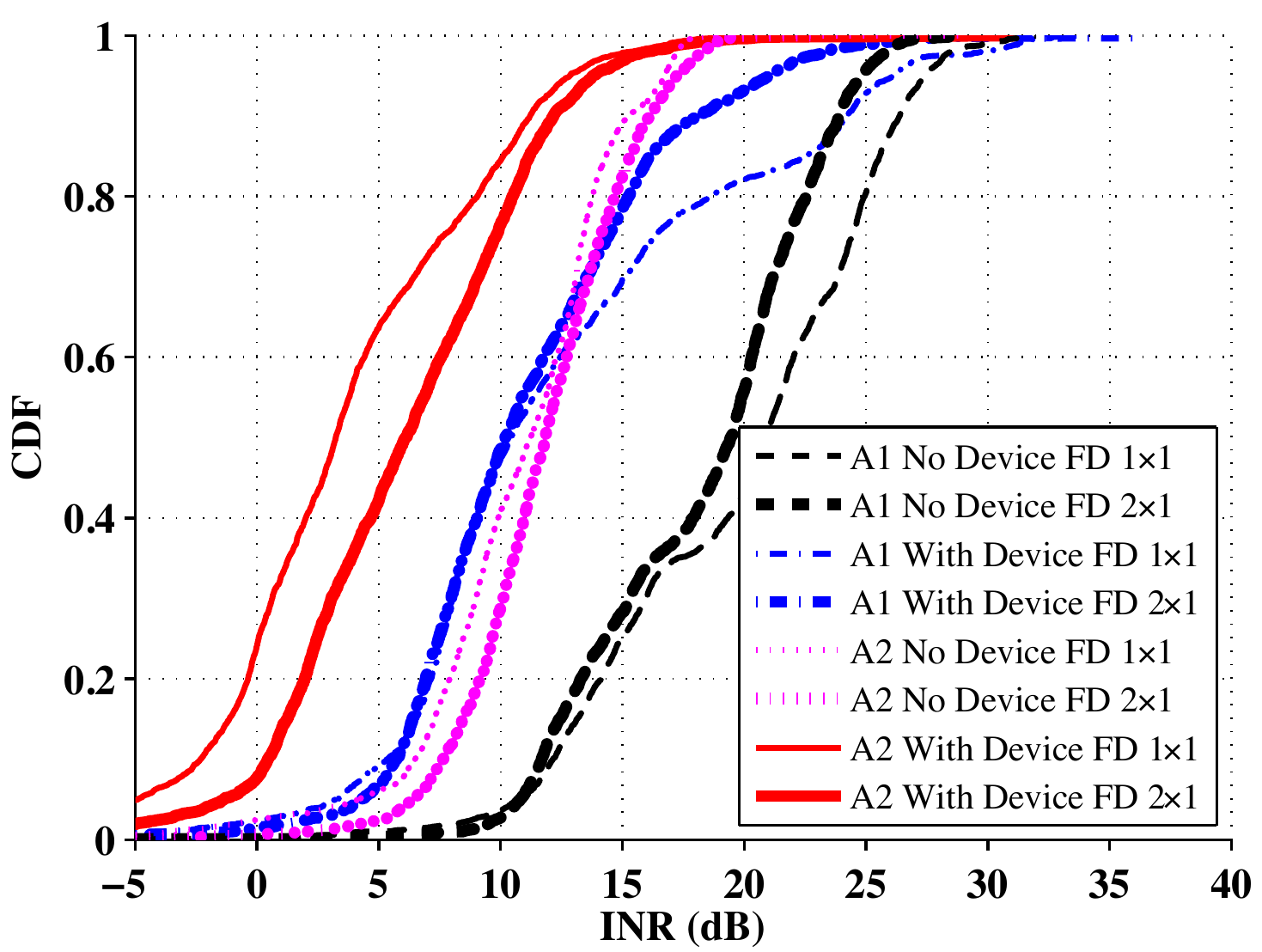}
   \caption{Residual self-interference INR (INR after all cancellation stages).}
   \label{fig:residualinr}
\end{figure}


\begin{figure*}[h]
\centering
\subfigure[Results for antennas placed around the device (device in the middle)]{
\includegraphics[width=0.48\textwidth]{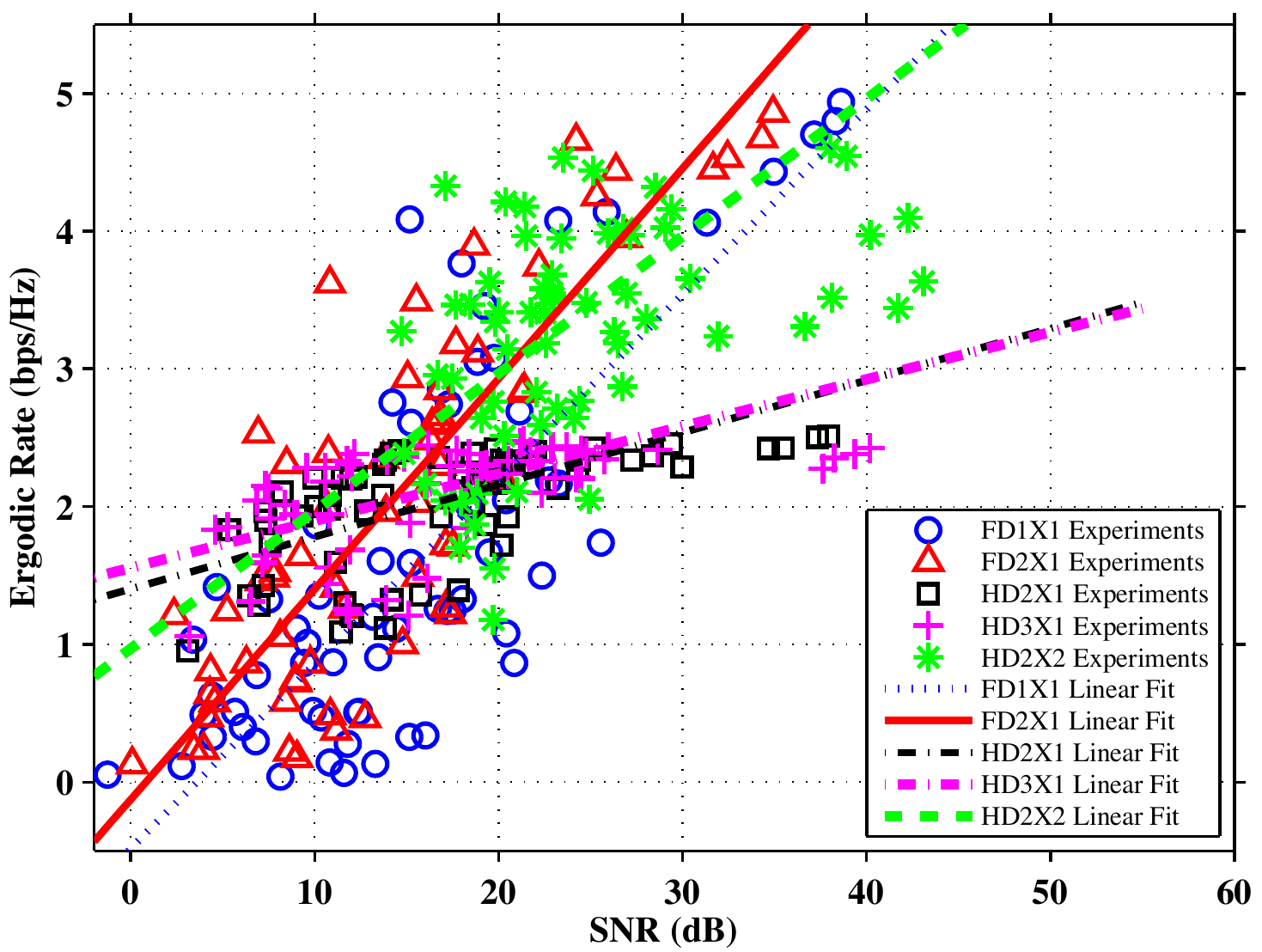}
   \label{fig:rates_withdevice}
}
\subfigure[Results for antennas without device in the middle.]{
\includegraphics[width=0.48\textwidth]{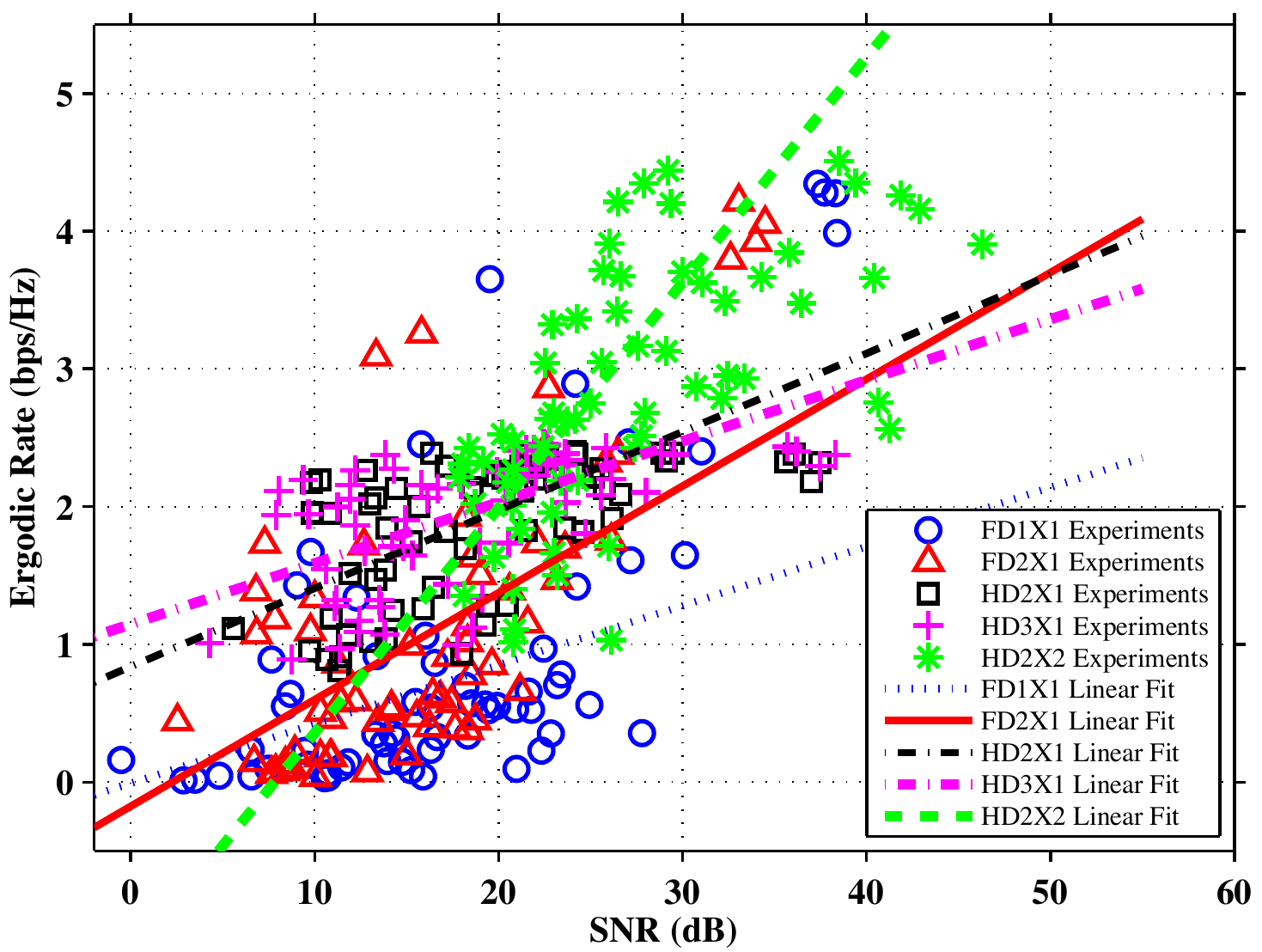}
   \label{fig:rates_nodevice}
}
\caption[]{Ergodic rate as a function of the SNR and and linear fit approximations.}
\label{fig:rates}
\end{figure*}

\cleardoublepage

\begin{figure}[h] 
   \centering
   \includegraphics[width=0.48\textwidth]{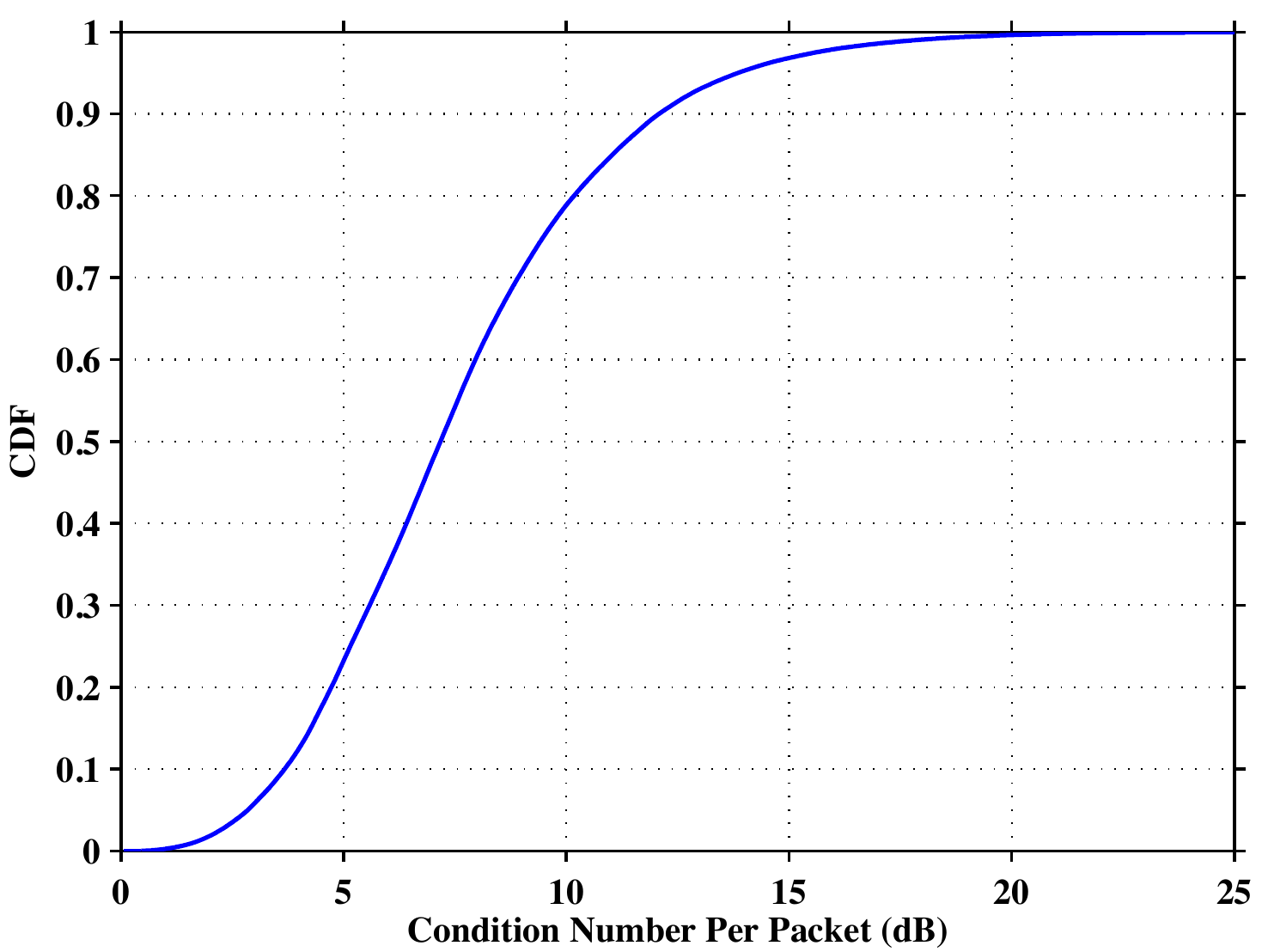}
   \caption{CDF of condition number per packet computed from experiment measeurements.}
   \label{fig:CNCDF}
\end{figure}

\begin{figure}[h] 
   \centering
   \includegraphics[width=0.48\textwidth]{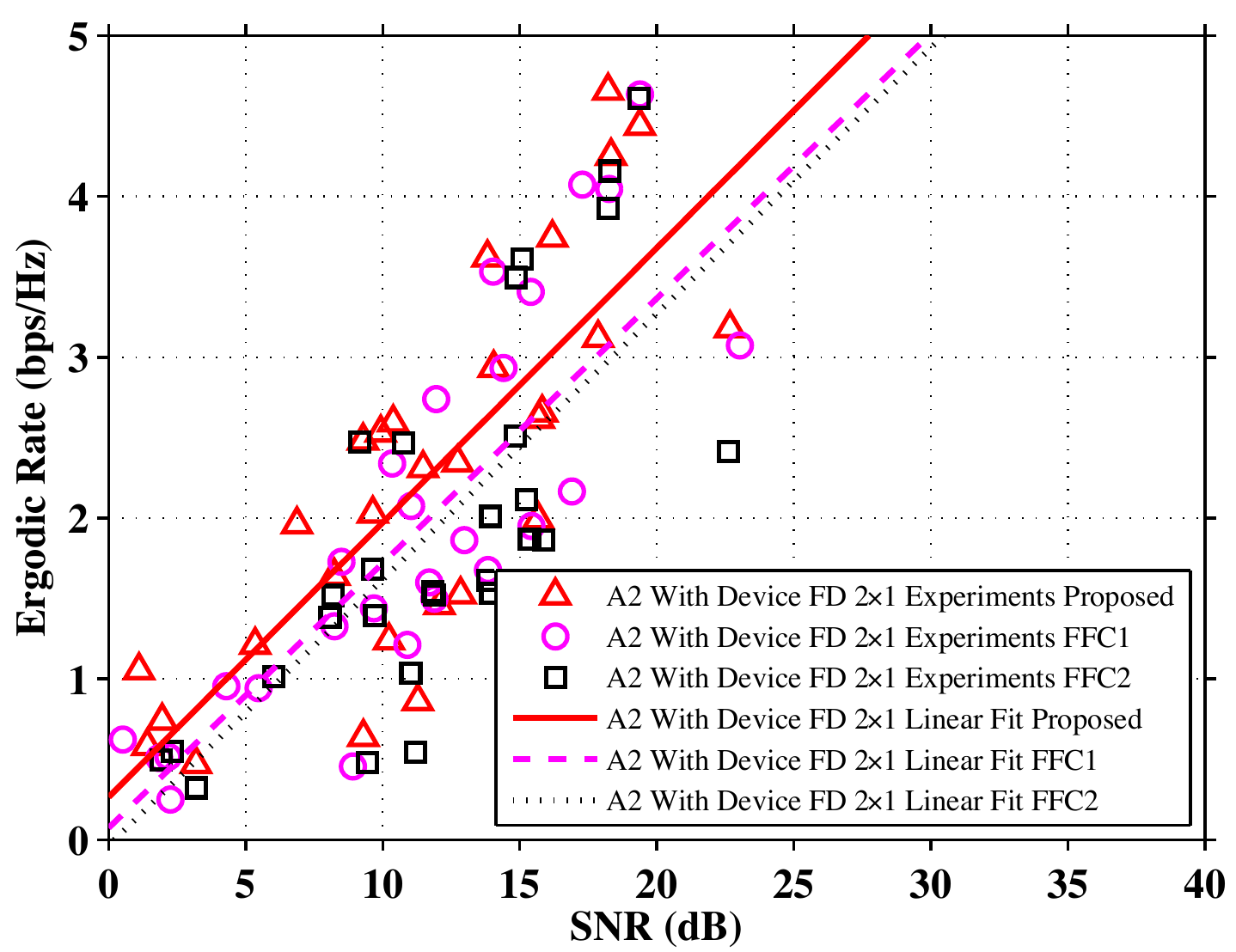}
   \caption{Ergodic rate vs. SNR performance for different analog cancelers}
   \label{fig:AR_active_freq}
\end{figure}

\begin{figure}[h] 
   \centering
   \includegraphics[trim=0.2in 7.1in 0.7in 1.45in, width=5in]{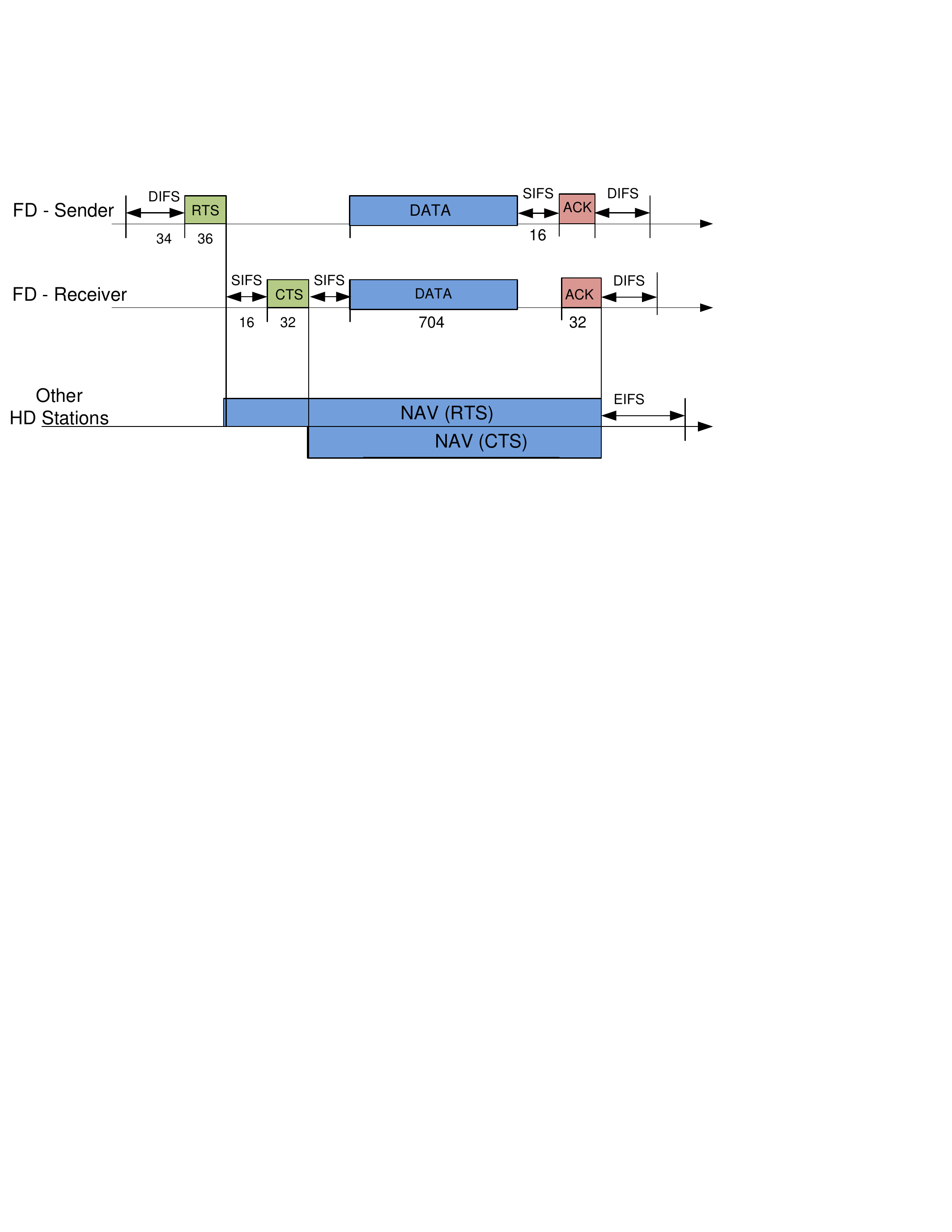}
   \caption{Full-duplex MAC frames for 802.11a}
   \label{mac}

\end{figure}


\begin{table}[h]
\small{
\centering
\begin{tabular}{l l}
\hline
Parameter & Value \\
\hline
\# Nodes & 2 (AP+1) to 9 (AP+8) \\
Transmit power & 9 dBm \\
AP to node distance & 14m \\
Center frequency & 2400 MHz \\
Free space path loss & 63 dB \\
RSSI & -59 dBm \\
Self-interf. cancellation & Default 85 dB \\
Modulation & QPSK, 18 Mbps\\
Data Asymmetry & 5\% to 100\% \\
Packet size & 40 to 1500 bytes uplink, 1500 bytes downlink \\
BER model & Q-function, treating self-interference as noise \\
Traffic model & Full buffer \\
AP Queue & Single queue \\
Buffer size &  25600 KB\\
Simulation time & 10 sec \\
\hline
\end{tabular}
\caption{MAC simulation parameter set}
\label{mactable}
}
\end{table}

\begin{figure}[h]
\centering
\includegraphics[trim=.63in 2.9in .87in 3.25in, clip, width=5in]{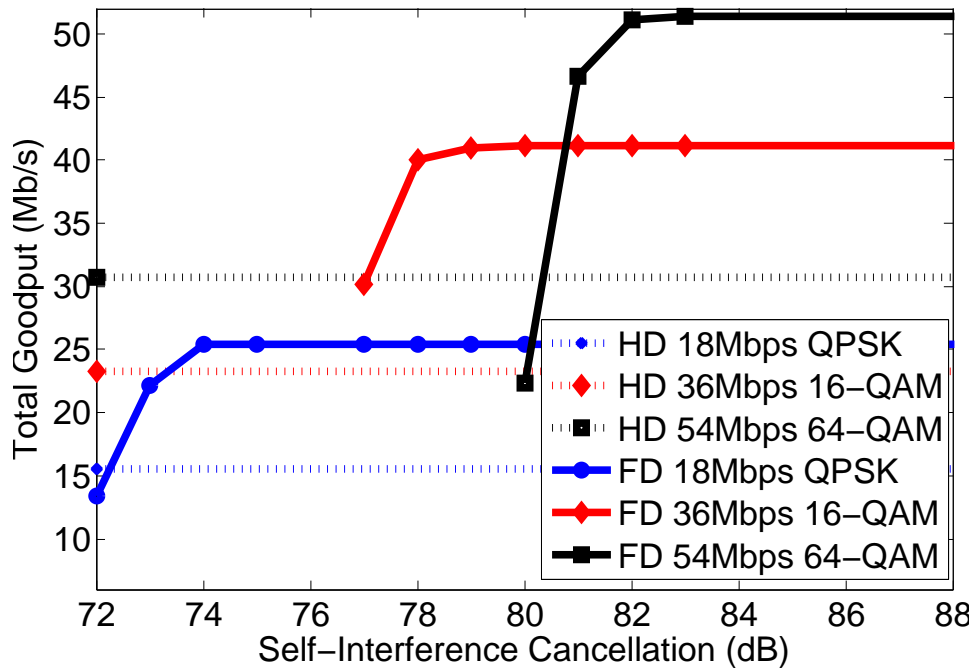}
   \caption{Performance of different modulations and constellations vs. self-interference cancellation, where half-duplex system does not use RTS/CTS signalling.}
   \label{newfig}
\end{figure}

\begin{table}[h]
\small{
\centering
\begin{tabular}{l l l l l l l l l}
\hline
 AP-Node & Node-AP &  HD w/o RTS  & HD with RTS    & FD       & Gain  & Gain\\
 (bytes) & (bytes) &  (Mbps)  & (Mbps)& (Mbps)   & (HD w/o rts)  & (HD with rts)    \\
\hline
1500 & 1500  & 6.92+6.77 &  6.41+6.39 & 12.81+12.81 & 1.87 & 2.00 \\
 1500 & 1000  & 7.85+5.26 & 7.21+4.92  & 12.80+8.54 & 1.62 &   1.76    \\
          1500 & 500    & 9.12+3.05 & 8.38+2.82  & 12.80+4.27 & 1.40 &  1.52     \\
 1500 & 40     & 10.83+0.3 & 9.81+0.26  & 12.80+0.34 & 1.18 &  1.30 \\
\hline
\end{tabular}
\caption{Goodput for varying packet sizes, AP + 1 STA, 85 dB cancellation, the goodput numbers are written as goodput from AP to station (downlink) + goodput from station to AP (uplink).}
\label{pktsize}
}
\end{table}

\begin{table}[h]
\centering
\begin{tabular}{|c|c|c|c|c|}
\hline
 &  Sum Goodput   &  \multicolumn{3}{|c|}{Average Goodput per station (Mbps)}    \\
 \hline
 &   (Mbps)  &   AP to HD STA  &   AP to FD STA/FD STA to AP  &   HD STA to AP \\
 &     &   (HD downlink)  &   (FD downlink=uplink)  &   (HD uplink) \\
\hline
$2m$ FD  &  2 & -  & $\frac{1}{n} $ &  -   \\
$2m$ HD   &  1  & $ \frac{1}{n(n+1)}$ & -  & $\frac{1}{n+1}$ \\
$m$ FD, $m$ HD  case 1 & $1+\frac{m}{n+1}$   & $\frac{1}{m(n+1)}$  & $\frac{1}{n+1}$  & $\frac{1}{n+1}$  \\
\hline
\end{tabular}
\caption{Theoretical normalized goodput for different scenarios, where number of nodes $n=2m$ and Case 1 is the case with modified half-duplex nodes.}
\label{tbl:anal}
\end{table}

\begin{table}[h]
\centering
\begin{tabular}{|c|c|c|c|c|}
\hline
 &  Sum Goodput   &  \multicolumn{2}{|c|}{Average Goodput (Mbps)}  & \% RTS/Data   \\
 \hline
 &   (Mbps)  &   AP to STA  &   STA to AP  &   Collisions \\
 &     &   (downlink)  &  (uplink)  &   \\

\hline
\hline
\multicolumn{5}{|c|}{$n=1$} \\
\hline
FD  &  25.62 &  12.81 & 12.81 & 10.8 \\
HD with RTS   &  12.8  & 6.41  & 6.39  & 11.1 \\
HD without RTS  &  13.69  & 6.92  & 6.77  & 10.7 \\
\hline
\hline
\multicolumn{5}{|c|}{$n=2$}\\
\hline
FD  &  26.02 & 6.52  & 6.52  &  8.1 \\
HD with RTS   &  12.82  &  2.12 & 4.29  & 17.8 \\
HD without RTS  & 13.25   & 2.13  & 4.49  & 17.8\\
\hline
\hline
\multicolumn{5}{|c|}{$n=4$}\\
\hline
FD  & 25.99  & 3.25  &  3.25 & 18.2\\
HD with RTS   & 12.78   & 0.67  &  2.53 & 26.9 \\
HD without RTS  &  12.57  &  0.63 & 2.49  &  26.7\\
\hline
\hline
\multicolumn{5}{|c|}{$n=8$}\\
\hline
FD  &  25.6 & 1.6  & 1.6  & 32.3 \\
HD with RTS   & 12.7   &  0.2  &  1.39  &  40.0 \\
HD without RTS  &  11.8  &  0.16  &  1.31  &  35.8 \\
\hline
\hline
\end{tabular}
\caption{Goodputs with multiple nodes.}
\label{tbl:multiplenodemac}
\end{table}

\begin{table}[h]
\centering
\begin{tabular}{|c|c|c|c|c|}
\hline
 &  Sum Goodput   &  \multicolumn{3}{|c|}{Average Goodput (Mbps)}    \\
 \hline
 &   (Mbps)  &   AP to HD STA  &   AP to FD STA/FD STA to AP  &   HD STA to AP \\
 &     &   (HD downlink)  &   (FD downlink=uplink)  &   (HD uplink) \\

\hline
\hline
\multicolumn{5}{|c|}{$m=1$}\\
\hline
$2m$ FD  &  26.07 & -  & 6.52  &  -   \\
$2m$ HD   &  12.82  &  2.12 & -  & 4.29 \\
$m$ FD, $m$ HD  case 1 & 17.87   & 3.81  & 4.81  & 4.45  \\
$m$ FD, $m$ HD case 2  & 18.64   & 4.79  & 5.65  &  2.55 \\
\hline
\hline
\multicolumn{5}{|c|}{$m=2$}\\
\hline
$2m$ FD  &  25.99 & -  & 3.25  &  -   \\
$2m$ HD   &  12.78  &  0.67 & -  & 2.53 \\
$m$ FD, $m$ HD  case 1  & 18.15   & 1.37 & 2.59 &  2.54 \\
$m$ FD, $m$ HD case 2 & 20.59   & 1.67  & 3.78  &  1.07 \\
\hline
\hline
\multicolumn{5}{|c|}{$m=4$}\\
\hline
$2m$ FD  &  25.6 & -  & 1.6  &  -   \\
$2m$ HD   &  12.7  &  0.2 & -  & 1.39 \\
$m$ FD, $m$ HD  case 1 & 18.37   & 0.5  & 1.38  & 1.32 \\
$m$ FD, $m$ HD case 2 & 21.86   & 0.59  & 2.23  &  0.42 \\
\hline
\hline
\end{tabular}
\caption{Goodputs with coexistence for multiple nodes where in case 1, HD nodes ignore collisions in NAV while in case 2, HD nodes are legacy and do not ignore collisions.}
\label{tbl:coexist_noeifs}
\end{table}

\begin{figure}[h] 
   \centering
   \includegraphics[trim=1.5in 1.5in 1.25in 0.4in, width=4in]{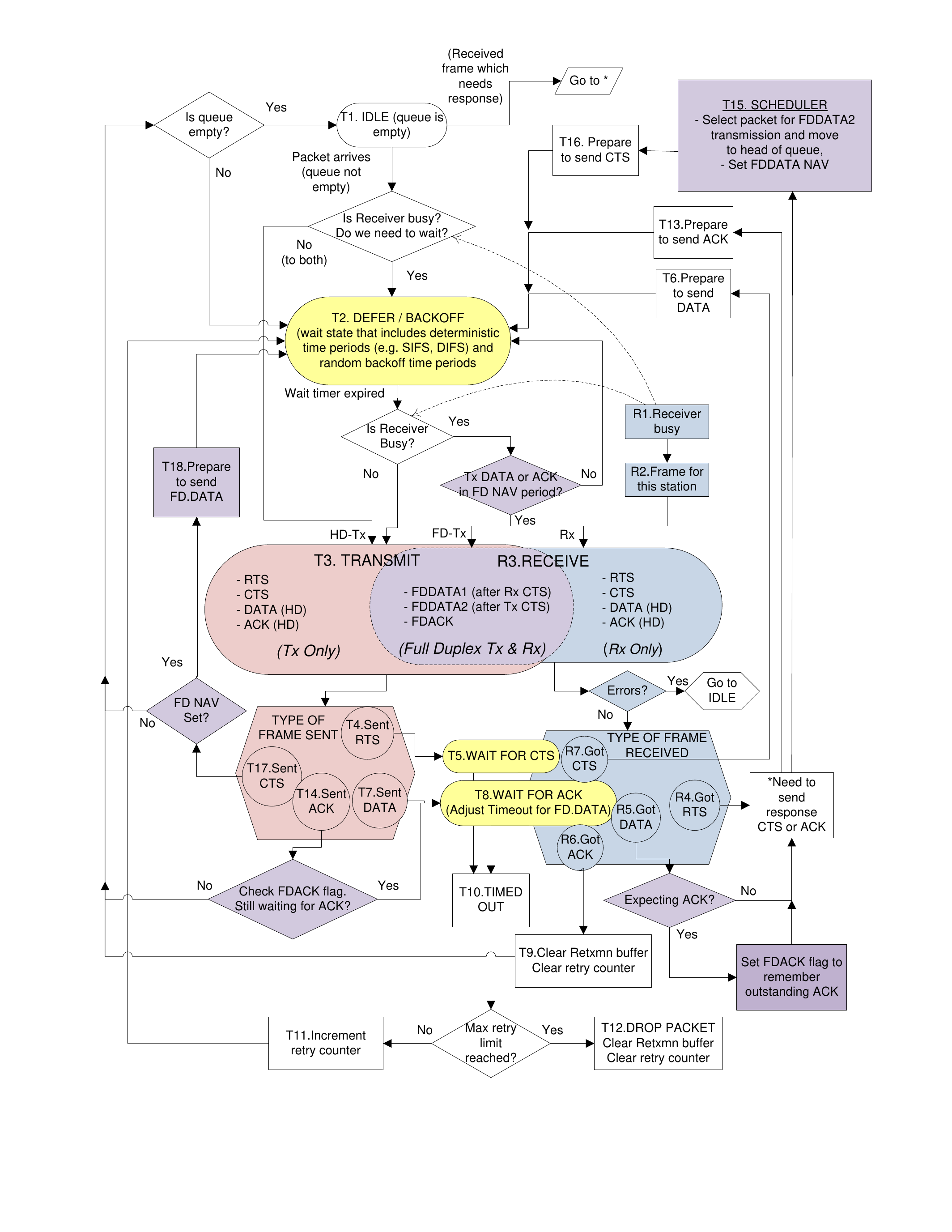}
   \caption{FD-MAC flowchart}
   \label{fdmac}
\end{figure} 